\documentclass[smallcondensed]{svjour3}     
\smartqed  

\RequirePackage{fix-cm}

\AtBeginDocument{%
  \paperwidth=\dimexpr
    1in + \oddsidemargin
    + \textwidth
    + 1in + \oddsidemargin
  \relax
  \paperheight=\dimexpr
    1in + \topmargin
    + \headheight + \headsep
    + \textheight
    + 1in + \topmargin
  \relax
  \usepackage[pass]{geometry}\relax
}
\usepackage{cite}
\usepackage{amsmath,amssymb,amsfonts}
\usepackage{algorithmic}
\usepackage{graphicx}
\usepackage{textcomp}
\usepackage{soul}	
\usepackage{balance}
\usepackage{tabularx}
\usepackage{enumitem}
\usepackage{adjustbox}
\usepackage{booktabs}
\usepackage{multirow}
\usepackage{caption}
\usepackage{dirtytalk} 
\usepackage{xcolor}
\usepackage{xcolor}
\usepackage[most]{tcolorbox}
\usepackage{pifont}
\usepackage[mathlines,switch]{lineno}

\usepackage{tikz}
\usepackage{graphicx}
\usepackage{listings}
\usepackage{pifont}
\usepackage{amsmath}
\usepackage{pdflscape} 
\usepackage{tikz}
\usepackage{dirtytalk}
\usepackage{makecell}
\usepackage[title]{appendix}
\usepackage{pythonhighlight}
\usepackage{caption}

\definecolor{LightGray}{gray}{0.9}
\usepackage[autostyle]{csquotes}
\usepackage{longtable}
\usepackage{booktabs}
\usepackage{tablefootnote}
\usepackage{multirow}
\usepackage{adjustbox}
\usepackage{enumitem}
\usepackage{siunitx}
\usepackage{fancyvrb}
\usepackage{soul}
\usepackage{textcomp}
\usepackage{tabularx}
\newlength\someheight
\usepackage{pgfplots}
\usepackage{pgfplotstable}
\pgfplotsset{compat=1.14}
\definecolor{storeClusterComponent}{HTML}{808080}
\definecolor{dbscan}{HTML}{BEBEBE}
\definecolor{constructCluster}{HTML}{DCDCDC}
\usepackage{tikz} 
\usepackage{subcaption}
\usetikzlibrary{shapes,calc,shadows,arrows}
\usepackage{pgf-pie}
\usepackage{etoolbox}
\newtoggle{showpct}
\usepackage{xcolor,colortbl}

\definecolor{codegreen}{rgb}{0,0.6,0}
\definecolor{codegray}{rgb}{0.5,0.5,0.5}
\definecolor{codepurple}{rgb}{0.58,0,0.82}
\definecolor{backcolour}{rgb}{0.95,0.95,0.92}

\lstdefinestyle{mystyle}{
    backgroundcolor=\color{backcolour},   
    commentstyle=\color{codegreen},
    keywordstyle=\color{magenta},
    numberstyle=\tiny\color{codegray},
    stringstyle=\color{codepurple},
    basicstyle=\ttfamily\footnotesize,
    breakatwhitespace=false,         
    breaklines=true,                 
    captionpos=b,                    
    keepspaces=true,                 
    numbers=left,                    
    numbersep=5pt,                  
    showspaces=false,                
    showstringspaces=false,
    showtabs=false,                  
    tabsize=2
}
\usepackage[most]{tcolorbox}

\newtcbtheorem{Summary}{\bfseries Summary}{enhanced,drop shadow={black!50!white},
  coltitle=black,
  top=0.3in,
  attach boxed title to top left=
  {xshift=1.5em,yshift=-\tcboxedtitleheight/2},
  boxed title style={size=small,colback=pink}
}{summary}

\newtcolorbox[auto counter]{summary}[1][]{title={\bfseries Summary~\thetcbcounter},enhanced,drop shadow={black!50!white},
  coltitle=black,
  top=0.3in,
  attach boxed title to top left=
  {xshift=1.5em,yshift=-\tcboxedtitleheight/2},
  boxed title style={size=small,colback=pink},#1}

\usepackage[round]{natbib}

\usepackage{etoolbox}
\makeatletter
\patchcmd{\pgfpie@slice}%
{\scalefont{#3}\beforenumber#3\afternumber}%
{\iftoggle{showpct}{\scalefont{#3}\beforenumber#3\afternumber}{}}%
{}{}

\patchcmd{\NAT@citex}
  {\@citea\NAT@hyper@{%
     \NAT@nmfmt{\NAT@nm}%
     \hyper@natlinkbreak{\NAT@aysep\NAT@spacechar}{\@citeb\@extra@b@citeb}%
     \NAT@date}}
  {\@citea\NAT@nmfmt{\NAT@nm}%
   \NAT@aysep\NAT@spacechar\NAT@hyper@{\NAT@date}}{}{}

\patchcmd{\NAT@citex}
  {\@citea\NAT@hyper@{%
     \NAT@nmfmt{\NAT@nm}%
     \hyper@natlinkbreak{\NAT@spacechar\NAT@@open\if*#1*\else#1\NAT@spacechar\fi}%
       {\@citeb\@extra@b@citeb}%
     \NAT@date}}
  {\@citea\NAT@nmfmt{\NAT@nm}%
   \NAT@spacechar\NAT@@open\if*#1*\else#1\NAT@spacechar\fi\NAT@hyper@{\NAT@date}}
  {}{}

\def\srcfile[#1,#2]#3{
    \node [draw, fill=white, minimum height=1.7cm, minimum width=1.35cm, rounded corners, double copy shadow={shadow xshift=0.1cm, shadow yshift=0.1cm}] (#1) at #2 {};
    \node[align=center, font=\sffamily\fontsize{5}{1.5}\selectfont] at #2 {#3};
}

\makeatother

\definecolor{dkgreen}{rgb}{0,0.6,0}
\definecolor{gray}{rgb}{0.5,0.5,0.5}
\definecolor{mauve}{rgb}{0.58,0,0.82}
\definecolor{dgreen}{rgb}{0.0, 0.5, 0.0}

\usepackage{mdframed}
\newcounter{finding}
\newmdenv[%
    linewidth=0.6pt,
    linecolor=black,
    outerlinewidth=0pt,
    skipabove=0pt,
    skipbelow=0pt,
    settings={\global\refstepcounter{finding}},
]{myfinding}

\makeatletter
\newcommand\notsotiny{\@setfontsize\notsotiny\@vipt\@viipt}
\makeatother

\usepackage{amsthm}

\theoremstyle{definition}

\usepackage{threeparttable}

\usepackage{array}
\definecolor{dodgerblue}{RGB}{30,144,255}
\definecolor{orange}{RGB}{255, 120, 8}

\newcommand{\lstbg}[3][0pt]{{\fboxsep#1\colorbox{#2}{\strut #3}}}
\lstdefinelanguage{diff}{
  basicstyle=\ttfamily\small,
  morecomment=[f][\lstbg{red!20}]-,
  morecomment=[f][\lstbg{green!20}]+,
  morecomment=[f][\textit]{@@},
}

\newcolumntype{C}[1]{>{\centering\arraybackslash}m{#1}}

\tcbset{commonstyle/.style={boxrule=0pt,sharp corners,enhanced jigsaw,nobeforeafter,boxsep=0pt,left=\fboxsep,right=\fboxsep}}
\newtcolorbox{mycolorbox}[1][]{commonstyle,#1}

\usepackage{svg}

\usepackage{etoolbox}
\newcommand*{\affaddr}[1]{#1} 
\newcommand*{\affmark}[1][*]{\textsuperscript{#1}}

\usepackage{scalerel}
\usepackage{tikz}
\usetikzlibrary{svg.path}

\definecolor{orcidlogocol}{HTML}{A6CE39}
\tikzset{
  orcidlogo/.pic={
    \fill[orcidlogocol] svg{M256,128c0,70.7-57.3,128-128,128C57.3,256,0,198.7,0,128C0,57.3,57.3,0,128,0C198.7,0,256,57.3,256,128z};
    \fill[white] svg{M86.3,186.2H70.9V79.1h15.4v48.4V186.2z}
                 svg{M108.9,79.1h41.6c39.6,0,57,28.3,57,53.6c0,27.5-21.5,53.6-56.8,53.6h-41.8V79.1z M124.3,172.4h24.5c34.9,0,42.9-26.5,42.9-39.7c0-21.5-13.7-39.7-43.7-39.7h-23.7V172.4z}
                 svg{M88.7,56.8c0,5.5-4.5,10.1-10.1,10.1c-5.6,0-10.1-4.6-10.1-10.1c0-5.6,4.5-10.1,10.1-10.1C84.2,46.7,88.7,51.3,88.7,56.8z};
  }
}

\newcommand\orcidicon[1]{\href{https://orcid.org/#1}{\mbox{\scalerel*{
\begin{tikzpicture}[yscale=-1,transform shape]
\pic{orcidlogo};
\end{tikzpicture}
}{|}}}}

\usepackage{listings}
\usepackage{xcolor} 

\lstdefinestyle{python}{
    language=Python,
    basicstyle=\ttfamily\small,
    keywordstyle=\color{blue}\bfseries,
    commentstyle=\color{green!50!black}\itshape,
    stringstyle=\color{red},
    showstringspaces=false,
    frame=single,
    numbers=left,
    numberstyle=\tiny\color{gray},
    stepnumber=1,
    numbersep=5pt,
    breaklines=true,
    breakatwhitespace=true,
    tabsize=4,
    captionpos=b
}

\usepackage[hidelinks, bookmarksnumbered=true]{hyperref}

\journalname{Empirical Software Engineering}
\begin{document}
\setcounter{tocdepth}{2}
\setcounter{secnumdepth}{3}

\title{An Empirical Study on Logging Evolution On Stack Overflow: Trends, Topics, and Challenges}

\author{Patrick Loic Foalem \and Andre Nguimbous \and Foutse Khomh \and Heng Li \and Ettore Merlo
}

\authorrunning{Patrick Loic Foalem\and Andre Nguimbous \and Foutse Khomh \and Heng Li \and Ettore Merlo}


\institute{ \affmark[*]Corresponding author. \\
\\
           Patrick Loic Foalem \and  Andre Nguimbous \and Foutse Khomh \and Heng Li \and  Ettore Merlo \at
              \affaddr{Department of Computer Engineering and Software Engineering \\ Polytechnique Montreal \\
              Montreal, QC, Canada} \\
              \email{\{patrick-loic.foalem \and  andre.nguimbous \and  foutse.khomh \and  heng.li \and ettore.merlo\}@polymtl.ca}           
}

\date{Received: date / Accepted: date}

\maketitle

\section*{Abstract}
\textbf{Context:} Logging is a crucial practice in software engineering, aiding developers in debugging applications when errors occur. While existing research has explored logging challenges from an academic perspective through literature reviews and source code analysis, a comprehensive study from the practitioners' perspective remains lacking. 

\noindent\textbf{Objective:} This paper aims to bridge this knowledge gap by presenting an in-depth analysis of trends, topics, and challenges in logging based on a dataset of 216,094 posts from Stack Overflow (SO), a popular Q\&A platform for developers, and complemented by a practitioner survey to validate the industrial relevance and difficulty of the identified topics. 

\noindent\textbf{Method:} 
We analyzed longitudinal trends by examining metadata related to users, questions, and tags associated with logging discussions. To identify prevalent discussion topics, we employed a Large Language Model (LLM)–based classification approach, based on a manually validated ground-truth sample. Topic popularity was assessed through average scores and views, while difficulty was measured using three community-driven metrics: the proportion of questions without accepted answers, the proportion of unanswered questions, and the median time to receive an accepted answer. To complement these findings, we conducted an industry-oriented survey in which 12 senior developers assessed the relevance, perceived difficulty, and engineering effort associated with the identified topic.

\noindent\textbf{Results:} 
Our analysis identifies 11 distinct topics, with the top three (General Logging Practices, Error Handling and Debugging, and Logging Levels and Output) accounting for over 70\% of all logging-related discussions. Notably, \textit{Logging in Containerized Environments} emerged as the most difficult topic: 64.9\% of its questions lack an accepted answer, and its median resolution time is among the highest. Survey results corroborate these findings, indicating that topics involving containerization, CI pipelines, and custom logging infrastructures are perceived by practitioners as particularly challenging in industrial settings. These findings highlight enduring practitioner struggles with logging in Docker or other containerized environments  and the integration of logging pipelines into orchestrators such as Kubernetes and cloud environments. 

\noindent\textbf{Conclusion:} This study sheds light on the practical challenges of logging and provides actionable insights for developers, framework vendors, researchers, and educators.


\section{Introduction}
\label{sec:introduction}

Logging is a fundamental practice in software engineering, crucial for recording dynamic information during a program's execution. This process supports real-time analysis and postmortem diagnostics, providing essential data for tasks such as failure diagnosis \citep{fu2013contextual}, performance analysis \citep{nagaraj2012structured}, and anomaly detection \citep{fu2009execution}. To facilitate logging, developers often employ specialized libraries such as Log4j and Logback, which are designed to integrate logging statements seamlessly into Java applications. These libraries allow for the definition of log messages that can describe runtime events using both static and dynamic data.

Each log entry is categorized by a verbosity level such as fatal, error, info, or debug to indicate the severity and importance of the event. For example, logs classified as `fatal' are critical and may halt the program after logging, whereas `debug' logs are more verbose and generally used during the development phase for detailed tracing. The choice of verbosity level helps developers manage the trade-off between obtaining comprehensive log data and minimizing performance overhead. In production environments, logging becomes particularly crucial as it may provide the only real-time data available to developers for diagnosing issues affecting end-users. Consequently, the strategic structuring and management of log data, from its generation to its archival, are vital for the effective maintenance and troubleshooting of software systems.

\begin{lstlisting}[language=Java, caption=Java Log4j logging example, label={lst:code1}]
import org.apache.log4j.Logger;
import org.apache.log4j.BasicConfigurator;

public class Log4jExample {
    /* Get actual class name to be printed on */
    static Logger log = Logger.getLogger(Log4jExample.class.getName());

    public static void main(String[] args) {
        // Configure logger
        BasicConfigurator.configure();
        int errorCount = 5; // Example of a dynamic variable
        // Log messages at various levels
        log.info("This is an info message");
        log.debug("This is a debug message");
        log.error("This is an error message with a dynamic variable: " + errorCount);
    }
}
\end{lstlisting}

Listing \ref{lst:code1} presents a Java code snippet that initializes the Log4j library and demonstrates logging at various severity levels. It includes an error message that incorporates a dynamic variable (`errorCount'), illustrating how log messages can be enriched with runtime values to enhance debugging capabilities.

Over the last decade, researchers have examined logging from multiple angles, including systematic reviews of research trends \citep{gholamian2021comprehensive, rong2017systematic, candido2021log, batoun2024literature}, empirical studies of industrial practices \citep{rong2020can, fu2014developers}, and qualitative investigations of developer perspectives \citep{li2020qualitative}. These studies provide valuable insights: they catalog the benefits and costs of logging, uncover challenges such as balancing log verbosity and overhead, and highlight inconsistencies between developer intentions and the resulting log statements in code.

Despite these contributions, significant gaps remain. Systematic literature reviews (SLR) and surveys synthesize knowledge across studies and participants, but they are either limited to the boundaries of prior academic work \citep{gholamian2021comprehensive, rong2017systematic, candido2021log, batoun2024literature} or a relatively small number of practitioners \citep{rong2020can, fu2014developers, li2020qualitative}. Industrial case studies, while rich in context, are typically constrained to a few organizations \citep{rong2020can, fu2014developers}. As a result, we lack a large-scale, cross-cutting view of daily logging challenges developers face across diverse contexts, technologies, and expertise levels.

To complement the literature, we explore developers' perspectives expressed on Q\&A platforms. Stack Overflow (SO)\footnote{\url{https://stackoverflow.com/}}, for example, hosts over 24 million questions and 36 million answers spanning a wide range of technologies, with 29 million registered developers as of August 2025. Investigating logging-related questions posted on Q\&A platforms such as Stack Overflow could reveal logging-related challenges faced by developers in their daily development activities, complementing traditional surveys and interview-based studies.

For example, one SO post\footnote{\url{https://stackoverflow.com/questions/2031163}} ask: \textit{``There are different ways to log messages, in order of fatality: FATAL; ERROR; WARN; INFO; DEBUG; TRACE; How do I decide when to use which?; What's a good heuristic to use?"}. This post received an answer within five minutes yet has accumulated over 867K views, suggesting that while the issue may be readily resolved, it continues to attract significant attention. Another example is this post\footnote{\url{https://stackoverflow.com/questions/43937031}} where developers faced issues configuring \textit{``logback"}, a logging framework, and received an accepted answer only after nine months despite being viewed over 25K by other developers; it may indicate that the problem encountered by this developer is difficult to solve.    

While we acknowledge the limitations of Stack Overflow, such as the lack of demographic and organizational metadata, its breadth and openness offer a unique vantage point on logging challenges as they emerge in everyday development practice. Methods such as surveys, interviews, and systematic literature reviews (SLRs) provide richer contextual and experiential insights, but are typically constrained by smaller sample sizes and limited coverage across technologies and ecosystems.
In this study, we leverage the complementary strengths of these approaches. Stack Overflow enables large-scale, longitudinal analysis of logging-related discussions, capturing recurring issues across programming languages, frameworks, and deployment contexts at a scale unattainable through traditional empirical methods. To mitigate its inherent limitations and ground our findings in industrial reality, we complement the Stack Overflow analysis with a targeted practitioner survey. The survey allows us to validate the industrial relevance, perceived difficulty, and engineering effort associated with the identified topics, thereby triangulating community-level signals with practitioners' lived experience.
Together, this mixed-methods design enables us to both map logging challenges at scale and assess how these challenges manifest in real-world development settings. By systematically analyzing 216,094 logging-related posts on Stack Overflow and validating the resulting topic taxonomy and difficulty patterns through practitioner feedback, this paper provides a more holistic and empirically grounded understanding of logging practices and challenges. To this end, we address the following research questions (RQs):



\noindent\hspace*{2mm} \textbf{RQ1 (Trend): How have logging discussions on Stack Overflow grown over the years?} This research question aims to comprehensively understand the global trends and technological contexts within which logging-related discussions occur on Stack Overflow. We analyzed a dataset comprising 216,094 logging-related questions, 118,196 responses, 146,840 distinct users, and 20,758 associated tags, collected from the Stack Exchange Data Dump\footnote{\url{https://archive.org/details/stackexchange}}
 by filtering (i) posts whose title contains the stem “log”, (ii) posts tagged with the stem “log”, and (iii) posts tagged with known logging-library tags. The library-tag list covers multiple ecosystems, e.g., Java/JVM: \texttt{log4j}, \texttt{log4j2}, \texttt{slf4j}, \texttt{logback}, \texttt{log4jdbc}; .NET: \texttt{log4net}, \texttt{nlog}, \texttt{serilog}; JavaScript/Node: \texttt{winston}, \texttt{bunyan}, \texttt{pino}, \texttt{pinojs}, \texttt{log4js}, \texttt{nest-winston}; Python: \texttt{structlog}, \texttt{loguru}; C/C++: \texttt{spdlog}, \texttt{glog}, \texttt{log4cxx}, \texttt{log4cpp}, \texttt{log4cplus}; Go: \texttt{logrus}, \texttt{zerolog}, \texttt{zap} (a.k.a. \texttt{go-zap}); Ruby/PHP: \texttt{log4r}, \texttt{monolog}; Mobile/Apple: \texttt{timber} (Android), \texttt{cocoalumberjack}; Log backends/observability: \texttt{grafana-loki}; MLOps: \texttt{mlflow}, \texttt{tensorboard}, \texttt{clearml}, \texttt{comet-ml}, \texttt{wandb}. The full list (40 tags) and extraction scripts are included in the replication package \citep{replication}. Our findings indicate that developers have relied heavily on SO to resolve logging issues, although participation patterns show that most users contribute only once, with limited sustained expertise across discussions.  
When compared with prior studies, logging discussions resemble Continuous Integration (CI) and refactoring in their fragmented engagement.
In terms of technological context, logging stands out as a cross-disciplinary topic. Through manual review of a significant sample of tags, we found that the majority of associated tags were related to \textit{programming concepts} (e.g., \textit{int}, \textit{exception}, \textit{biginteger}) and \textit{frameworks/libraries} (e.g., \textit{Symfony}, \textit{JavaFX}, \textit{Vue.js}), highlighting its foundational role in software engineering. Unlike CI, where tags are dominated by platforms and servers, and unlike refactoring, where tags center on languages and algorithms, logging discussions are spread across multiple dimensions--tools, platforms, frameworks, and concepts. Notably, emerging domains such as machine learning (ML) (\textit{data-science}, \textit{text-classification}) and security (\textit{security}, \textit{protected}, \textit{credential-manager}) increasingly appear in logging contexts, reflecting the growing importance of observability and trustworthy system design.  
\\

\hspace*{2mm} \textbf{RQ2 (Topic): What topics are discussed around logging?} The goal of this research question is to identify the key topics that developers discuss on Stack Overflow related to logging, as well as to determine their relative frequency. To achieve this, we employed an LLM-based classification approach, grounded in a manually validated ground-truth sample. Our analysis reveals 11 distinct topics, with \textit{General Logging Practices (36.53\%)}, \textit{Error Handling and Debugging (21.27\%)}, and \textit{Logging Levels and Output (14.54\%)} accounting for more than 70\% of all logging discussions. Other specialized topics, such as \textit{Data and TensorBoard Logging (2.56\%)}, \textit{Logging in Containerized Environments (2.36\%)}, and \textit{Testing and CI Pipelines (0.71\%)}, represent smaller but non-negligible portions of the discussion space, often tied to specific ecosystems or deployment contexts. To assess whether the topics identified through Stack Overflow mining reflect concerns that practitioners actually encounter in real-world development settings, we complement this analysis with an industry-oriented survey. In this survey, practitioners were asked to assess the \emph{industrial relevance} of each topic, defined as the extent to which the topic frequently and meaningfully appears in their day-to-day development, operation, and maintenance activities. The survey results confirm that the topic taxonomy derived from Stack Overflow captures logging concerns that are widely recognized in practice, including both broadly applicable topics and ecosystem- or context-specific ones. This complementary validation strengthens the relevance of the identified topics and supports their use as a foundation for subsequent analyses focusing on topic difficulty and associated challenges.\\
\hspace*{2mm} \textbf{RQ3 (Challenges): Which topics are the most popular and difficult among logging questions?} We leverage information from SO to identify the most popular and challenging topics in logging. To assess topic popularity and difficulty, we measured average scores and views as indicators of community attention, and difficulty through three established metrics: the proportion of questions without accepted answers, the proportion of unanswered questions, and the median time to receive an accepted answer. Our findings highlight sharp contrasts across topics. Logging in \textit{Containerized Environments} emerges as the most challenging, with 64.9\% of its questions lacking accepted answers and one of the highest median resolution times. These results underscore enduring struggles with the integration of logging into orchestrators such as Kubernetes, Docker, and cloud platforms. By contrast, \textit{General Logging Practices}--which covers fundamental concerns such as when, where, and how to insert log statements across different programming languages and frameworks--appears among the least difficult (51.56\% without accepted answers) and is resolved most quickly (median 21.36 hours). Similarly, \textit{Logging Levels and Output}--which deals with selecting appropriate severity levels (DEBUG, INFO, WARN, ERROR) and composing meaningful log messages--stands out as both highly popular and relatively easy to resolve. 
To complement these community-level difficulty signals, we conducted an industry-oriented survey to capture practitioners' perceptions of \emph{challenge} and \emph{engineering effort} associated with logging topics in real-world settings. While the survey does not aim to quantify difficulty at scale, it provides qualitative and perceptual insights into why certain topics are experienced as more challenging in practice. Survey responses corroborate the Stack Overflow findings, indicating that topics involving containerization, CI pipelines, and custom logging infrastructures require disproportionate engineering effort due to their cross-cutting nature, environmental dependencies, and operational constraints. Practitioners emphasized that logging effort increases substantially when it must support distributed execution, observability, security, and cost considerations, even when the underlying logging APIs appear simple.
Conversely, foundational topics such as \textit{General Logging Practices} and \textit{Logging Levels and Output} are perceived as requiring comparatively less engineering effort, as they benefit from well-established conventions, stable tooling, and abundant documentation. Interestingly, \textit{Data and TensorBoard Logging}, although among the least popular topics on Stack Overflow, was perceived by practitioners as effort-intensive when it arises, reflecting the additional complexity introduced by large data volumes, experiment tracking requirements, and storage and performance constraints in ML pipelines.
Together, the combination of Stack Overflow--based difficulty metrics and practitioner-reported effort provides a nuanced view of logging challenges. It reveals that logging difficulty is not solely a function of API complexity, but rather emerges from the interaction between logging mechanisms, system architecture, deployment environment, and organizational practices. Compared to related domains such as Continuous Integration and Refactoring, logging stands out as both more difficult and more time-consuming, with 56.49\% of questions lacking accepted answers and a median resolution time of 45.37 hours \citep{ouni2023empirical, peruma2022refactor}, a finding that is further reinforced by practitioners' accounts of sustained engineering effort in complex logging scenarios.


\noindent The primary contributions of this study are summarized as follows:
\begin{itemize}
    \item We conduct a comprehensive empirical study (mining and analyzing 216,094 SO posts related to logging) aiming to deepen the understanding of challenges, trends, and topics related to logging. 
    \item We employ a mixed-method approach, incorporating both quantitative and qualitative analyses, to elucidate the characteristics of logging-related topics and the utilization of logging frameworks.
    \item We complement the large-scale Stack Overflow analysis with an industry-oriented survey, enabling validation of the industrial relevance, perceived difficulty, and engineering effort associated with the identified logging topics from practitioners' perspectives.
    \item We compare our findings with related empirical studies on Continuous Integration (CI) \citep{ouni2023empirical} and Refactoring \citep{peruma2022refactor}, highlighting both similarities and differences across domains to position logging within the broader SE research landscape.
    \item We offer practical insights for researchers, developers, tool builders, and educators based on our findings.
    \item We provide a publicly accessible replication package \citep{replication}, which includes the dataset and scripts necessary for replicating our results.
\end{itemize}

The structure of this paper is as follows: Section \ref{sec:approach} outlines our research methodology. Section \ref{sec:result} presents our results and findings. Section \ref{sec: discussion_implication} discusses the implications of our study. Section \ref{sec:related_work} reviews related work. Section \ref{sec:threats_to_validity} addresses potential threats to validity. Finally, Section \ref{sec:conclusion} concludes the paper.

\section{Experiment setup} 
\label{sec:approach}

To explore logging practices and their evolution on Stack Overflow, we adopted a mixed-methods approach combining qualitative and quantitative analyses similar to methodologies used in prior studies ~\citep{gujral2018empirical,openja2020analysis,wen2021empirical,yahmed2023deploying,ouni2023empirical}. This approach focuses on collecting and analyzing posts from Stack Overflow that are relevant to logging. Our methodology, which is clearly illustrated in Figure 2, consists of two main steps: (1) extraction of logging posts, and (2) analysis of these posts.

\subsection{Extracting logging posts}
\begin{figure}
    \centering
    \includegraphics[width=\textwidth]{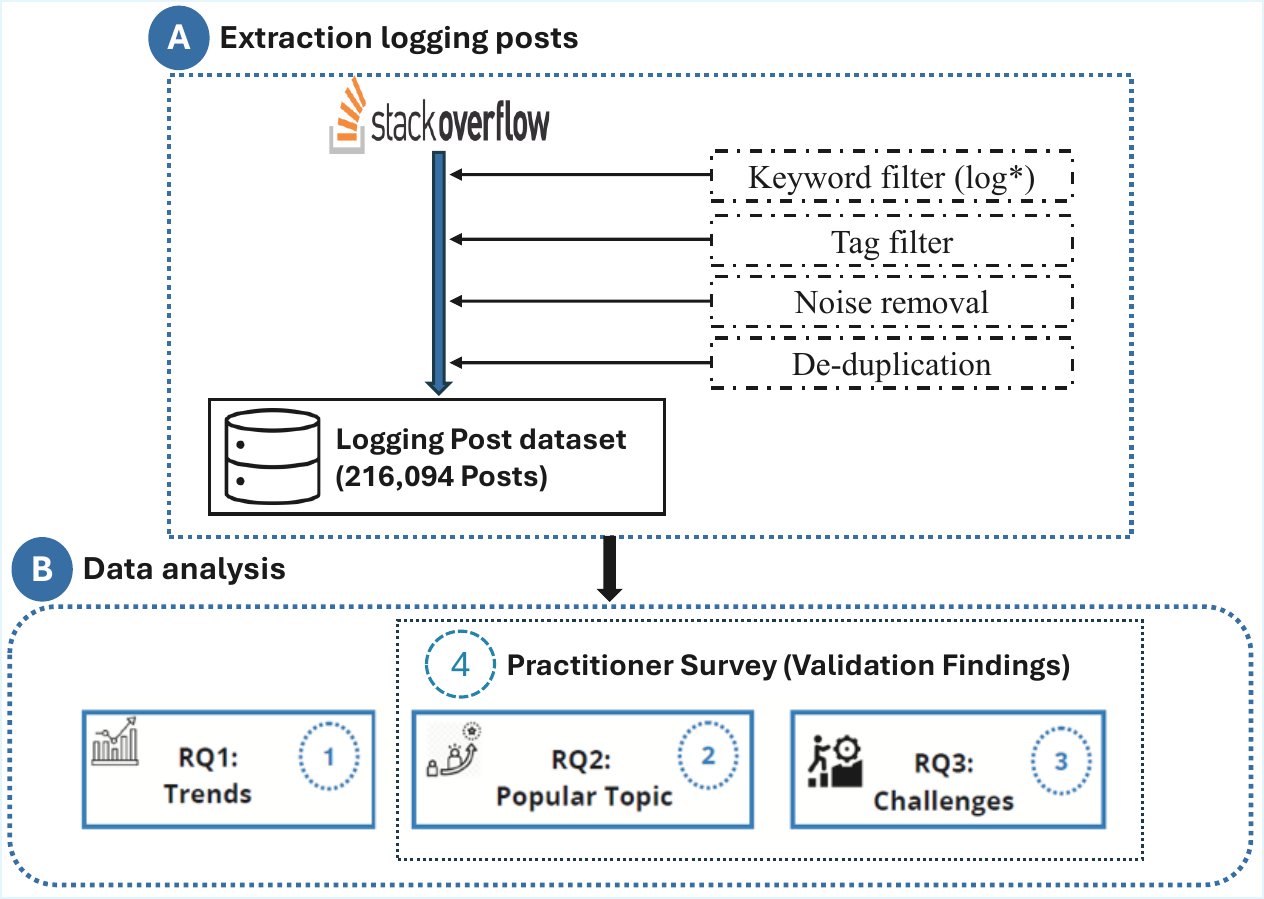}
    \caption{Overview of our methodology}
    \label{fig:methodology}
\end{figure}
\begin{itemize}[leftmargin=0.8 pt,align=left]
    \item[] \textbf{Stack Overflow Database:} The Stack Overflow dataset was obtained from the Stack Exchange Data Dump on June 6, 2025. This dataset contains posts dating back to July 31, 2008, and each post's metadata includes its identifier, type (e.g., Question, Answer, Wiki), creation date, body, title, score, view count, and up to five tags representing its topics. Notably, Stack Overflow allows the author of a question to validate an answer by marking it as ``accepted'', signifying its effectiveness in addressing the question. In the SO gamification model, users are encouraged to participate actively and provide quality content through a system of rewards and reputation points. Users can vote on the usefulness of posts, with upvotes increasing a post's visibility and the poster's reputation score. High-scoring contributions are more prominently displayed, stimulating users to produce accurate and helpful answers. This system enables a self-regulating community where expertise is recognized and rewarded, promoting an environment rich in professional peer-to-peer engagement and knowledge exchange.

    \item[] \textbf{Collect relevant logging posts}:
    To extract posts related to logging, we queried the Stack Exchange Data Dump using the stem ``log'' in both the \emph{Title} and \emph{Tags} fields (we excluded bodies to reduce false positives, since titles and tags succinctly convey a question's primary intent \citep{ouni2023empirical, rosen2016mobile, peruma2022refactor}). This yielded 387{,}412 title matches and 237{,}613 tag matches; after de-duplication our initial corpus was $D_0=\,$432{,}896 unique posts with $T_0=\,$16{,}266 distinct tags.
    
    Because broad ``log*'' queries admit many domain-external uses, we applied a reproducible sanitization pipeline to \emph{Title} and \emph{Tags} using a curated set of regular expressions to remove: authentication/login (\texttt{login}, \texttt{logout}, \texttt{authlogic}), UI dialogs (\texttt{*dialog*}, \texttt{messagebox}, etc), mathematics/statistics (\texttt{log-likelihood}, \texttt{logistic-regression}, \texttt{n log n}, \texttt{logsumexp}, \texttt{lognormal}, \texttt{logloss}, \texttt{logspace}, \texttt{logprob}, etc), programming languages/tools (\texttt{prolog}, \texttt{verilog}, \texttt{systemverilog}, \texttt{netlogo}, \texttt{anylogic}, \texttt{logisim}, \texttt{logitech}, \texttt{saleslogix}, etc), VCS commands (\texttt{git log}, \texttt{hg log}, \texttt{svn log}, etc), and tracking (\texttt{worklog}, \texttt{keylogger}, etc). The full list and code are provided in our replication package \citep{replication}. After filtering, the cleaned seed set was $D_0=\,$114{,}678 posts with $T_0=\,$16{,}100 tags.
    
    The first three authors of this paper, each with at least six years of experience in software engineering, artificial intelligence, and cloud computing, then manually reviewed $T_0$ to enumerate library-specific logging tags (some posts use library tags without the explicit word ``logging''; e.g., PostID~=~74989141\footnote{\url{https://stackoverflow.com/questions/74989141}}). We identified 40 logging libraries (listed in the replication package) and, by incorporating these tags, collected an additional $D_1=\,$97{,}209 posts.
    
    During auditing, we found a small number of posts initially removed by our negative patterns that were in fact about logging (e.g., PostID~=~4370282\footnote{\url{https://stackoverflow.com/questions/4370282}} matches the token \texttt{blog} but discusses logging). For such cases, we applied a whitelist promotion based on the presence of library-specific logging tags in $D_1$ and promoted true positives back into the seed. This increased the cleaned seed from 114{,}678 to 118{,}885 posts; we refer to this promoted seed as our final $D_0$.
    
    Finally, we merged the promoted seed $D_0$ with the library-expanded set $D_1$ and removed duplicates, yielding $D_2=\,$216{,}094 unique posts and $T_0 =\,$ 20,758 unique tags. Our complete dataset and scripts are available in the replication package \citep{replication}. Table~\ref{tab:collected_data} summarizes that, of the 216{,}094 posts, 118{,}196 (\textbf{54.7}\%) received at least one answer and 87{,}197 (\textbf{40.4}\%) have an accepted answer.
\begin{table}[htbp]
\centering
\caption{Overview of our collected dataset.}
\label{tab:collected_data}
\begin{tabular}{@{}lr@{}}
\toprule
Metric                             & Value \\ \midrule
Number of posts                  & 216,094 \\
Number of answered posts     & 118,196 \\
Number of accepted answers       & 87,197 \\
Number of distinct tags          & 20,758 \\
Number of distinct users         & 146,840 \\
Average number of tags per question & 3 \\
Average number of answers per question & 1.35 \\ \bottomrule
\end{tabular}
\end{table}
    
\end{itemize}

\subsection{Data analysis}
\begin{itemize}[leftmargin=0.8 pt,align=left]
    \item[] \textbf{Step 1. Trend (RQ1):} To assess the trend of logging on SO, we adapted methodologies from prior studies \citep{chen2020comprehensive,gujral2018empirical,openja2020analysis,lopez2018investigation,bagherzadeh2019going,rosen2016mobile,ouni2023empirical}.
    We first analyzed the following metrics: the number of questions, which measures the total number of logging-related queries posted annually, indicating the popularity and community interest over time; the number of questions with accepted answers, assessing the effectiveness of community responses by quantifying how many queries receive satisfactory resolutions each year; the number of questions without accepted answers, highlighting unresolved issues and potential knowledge gaps; and user engagement per year, reflecting the level of community involvement in logging discussions. Collectively, these indicators provide a comprehensive overview of the topic's trends on Stack Overflow \citep{chen2020comprehensive,wen2021empirical}. Subsequently, we analyzed statistics of SO users engaging in logging discussions to identify the subset of users responsible for logging questions and answers. Finally, we manually analyzed a statistically significant sample of the initial set of tags ($T_0 = 20{,}758$ tags) to identify the concepts and technologies associated with logging. To ensure the representativeness of our manual review, we selected a sample size corresponding to a 99\% confidence level and a 5\% confidence interval. The required sample size $N$ was computed using the standard formula for estimating proportions in finite populations:
    \[
    N = \frac{Z^2 \cdot p \cdot (1 - p)}{e^2}
    \]
    where $Z$ is the z-score associated with the desired confidence level ($Z = 2.58$ for 99\%), $p$ is the estimated proportion of the population (set conservatively to $0.5$ to maximize variance), and $e$ is the margin of error ($e = 0.05$). Applying this formula yields a minimum sample size of $N = 665$ tags, which we manually analyzed to construct the set of logging-related concepts and technologies \citep{SampleSize}. The first two authors independently coded the selected tags, made annotation notes, and then discussed their observations to resolve disagreements and converge on a finalized set of categories. This coding approach, which has been successfully applied in prior empirical studies of developer discussions on SO \citep{peruma2022refactor, ouni2023empirical}, allowed us to identify and group concepts or technologies associated with logging. The results of this first step are presented in Section 3.1 to address RQ1.
    
    \item[] \textbf{Step 2. Popular topic (RQ2):} The goal of this step is to extract prevalent topics from SO discussions related to logging practices. We employed a Large Language Model to assist in the discovery of these topics. Several studies have leveraged LLMs as evaluators or ``judges" in software engineering tasks, such as code generation, summarization, evaluation, and bug detection \citep{tong2024codejudge, he2025code, crupi2025effectiveness}. In this study, we utilize LLMs as a judge for topic discovery, similar to previous work in the field of software engineering \citep{mu2024large, doi2024topic, li2025software}. 

    We adopted a reference-based judging approach, as suggested by \citep{lin2024engineering}, which requires ground truth data from real-world sources to assess how well a sample aligns with an ideal response. For our LLM, we chose the GPT-4o-mini model, as it is cost-effective and provides performance comparable to GPT-4, while being more affordable for large-scale use \citep{verdet2023exploring}. We set the temperature parameter to 0, as done in previous studies \citep{verdet2023exploring}, to ensure deterministic and consistent outputs.
    
    \textbf{LLM-Based Topic Classification:}
    Using a fixed prompt, we asked the LLM to act as an expert in logging practices and software engineering. This approach optimized the LLM's ability to generate precise responses \citep{wang2024ai}. We prompted the LLM to classify Stack Overflow posts (both the title and body) into predefined logging topics. If no topic matched, the LLM was instructed to propose a new topic. Additionally, for each topic classification, the LLM was asked to extract important keywords to justify its classification, and the results were returned in JSON format (see Listing~\ref{lst:llm_topic_prompt} for the exact prompt).
    \begin{lstlisting}[language=Python, caption={Prompt used for LLM-based topic classification of Stack Overflow logging posts}, label={lst:llm_topic_prompt}]
    PROMPT = """You are an expert in logging practices and software engineering.
    Classify the following Stack Overflow post (already preprocessed as Title/Body text) into one of the predefined topics.
    If no topic matches well, propose a new topic name.
    Also extract 3-7 important keywords that justify your choice.
    
    Return JSON only with keys: "topic" and "keywords".
    
    Available Topics and Descriptions:
    1. Logging Levels and Output – Problems related to choosing log levels (INFO, DEBUG, ERROR, etc.), formatting log messages, and configuring log outputs.
    2. General Logging Practices – Questions about the general use of logging across languages and platforms (Python, Java, .NET, Android, etc.).
    3. Error Handling and Debugging – Issues where logging is connected to debugging errors, crashes, or error traces.
    4. File Logging and Configuration – Managing, configuring, and troubleshooting log files (rotation, paths, archiving).
    5. Custom Logging Frameworks – Building/extending custom logging systems, or integrating specialized logging into frameworks.
    6. Data and Tensorboard Logging – Logging in data/ML contexts (TensorBoard, MLflow, DVC, model metrics).
    7. Logback and Application Configuration – Configuring/troubleshooting Java Logback (often Spring Boot).
    8. Logging in Containerized Environments – Logging with Docker, Kubernetes, container runtimes and collectors.
    9. NLog and .NET Logging – Using or configuring NLog in .NET applications.
    10. Testing and CI Pipelines – Logging within automated tests or CI/CD (Selenium, Jest, Azure DevOps, etc.).
    11. Event Logging and Monitoring – Event logs for monitoring/observability/fault diagnosis (agents, event IDs, log analytics).
    
    Output strictly valid JSON.
    
    Post:
    
    {text}
    
    """
    \end{lstlisting}
    \textbf{Data Preprocessing:}
    Before prompting the LLM, we preprocessed the input (title + body) by removing HTML code, tags, and email addresses to avoid noise and ensure clean data.
    
    \textbf{Ground Truth Construction:}
    To build the ground truth sample for topic classification, we manually analyzed 10\% of our $D_2$ dataset. The goal was to identify an initial set of topics for the classification task. The first two authors, both PhD students with over six years of experience in software engineering, collaboratively analyzed the pilot sample through discussions and open coding. Open coding is a popular method used in software engineering to develop taxonomies \citep{verdet2023exploring, abbassi2025unveiling}. 
    During this analysis, the authors read through the titles and bodies of the posts and identified important keywords that could serve as initial topic labels. Without an initial set of logging topics, inefficient labels were iteratively refined and merged. This process led to the identification of 11 topics, which were used as the ground truth and starting point for classification. To reduce the risk of hallucination by the LLM \citep{sollenberger2024llm4vv}, we provided a brief description of each of the 11 topics in the prompt.

    \textbf{Topic Classification and Iteration:}
    The LLM was then asked to classify 40\% of the remaining $D_2$ dataset into the 11 predefined topics. This subset was deliberately selected for \emph{topic discovery}, with the explicit goal of assessing whether the LLM could identify previously unseen logging topics. For posts that did not clearly fit any existing topic, the LLM was instructed to propose new topic labels. 
    All LLM-generated classifications and proposed topics were jointly reviewed by the first two authors during dedicated meetings. In this phase, 85\% of the LLM-suggested topic labels were confirmed by the human reviewers without modification, indicating a high level of alignment between the LLM and expert judgment. All remaining cases of disagreement or misclassification were discussed collaboratively and reassigned to the most appropriate topic by consensus.
    This review revealed that the newly proposed topics were overly specific and largely confined to .NET framework logging practices. As these posts shared substantial semantic overlap with existing categories, they were systematically merged into the corresponding predefined topic, \textit{``NLog and .NET Logging.''} This process confirmed topic saturation and validated the completeness of the original topic set.

    \textbf{Final Manual Review and Consistency Assessment:}
    In the final step, the first two authors separately labelled the remaining 50\% of the dataset. We reserved this 50\% as a human-only holdout to ensure an independent reliability check. Once the entire dataset was labeled, we assessed the inter-rater reliability by calculating the Cohen's Kappa score \citep{mchugh2012interrater} to measure the consistency between the two reviewers. We achieved a Kappa score of 0.781, which aligns with the results from previous studies in software engineering \citep{foalem2024studying, abbassi2025unveiling}. All disagreement cases were resolved during the meeting discussion.    
    
    \item[] \textbf{Step 3. Challenges (RQ3):} To assess popular and difficult logging topics on SO, we utilize popular metrics from previous studies: (1) view count, and (2) score of logging posts to determine the most popular, high scores obtained for these metrics; the most popular is the logging topic. These metrics are then aggregated by topic, calculating the average score and average view count for all questions within each specific topic. This aggregation provides a comprehensive view of which topics are not only popular but also actively engage the community \citep{rosen2016mobile, alomar2020exploratory, peruma2022refactor}. We further evaluate difficult logging posts by examining questions that lack an ``accepted" answer and by analyzing the median time it takes to receive an accepted answer. This approach allows us to identify not only the most popular topics but also those that pose significant challenges to the community, reflecting both unresolved queries and longer wait times for solutions. Our methodology for identifying challenging topics follows established practices from prior research that analyze domain-specific challenges through Stack Overflow data, ensuring that our findings are grounded in a proven analytical framework \citep{openja2020analysis, peruma2022refactor, ouni2023empirical}. This step 3 helps us to answer our RQ3 discussed in section \ref{sec:RQ3}.

    \item[] \textbf{Step 4. Practitioner Survey:}

    To complement our large-scale analysis of logging-related discussions on Stack Overflow and to validate the findings associated with \textbf{RQ2--RQ3}, we designed an industry-focused survey targeting experienced software engineers. The objectives of this survey are twofold: (i) to assess whether the logging topics and broad difficulty trends identified through Stack Overflow align with practitioners' perceptions and industrial experience,  and (ii) to gain deeper qualitative insights into the factors that practitioners perceive as contributing to logging difficulty and engineering effort in practice. Importantly, the survey is intended to capture \emph{perceived} relevance, difficulty, and effort rather than to establish an objective measure of logging complexity.
    Survey-based empirical studies are a well-established method in software engineering research for capturing practitioners' perceptions and contextualized experience, and they effectively complement mining-based analyses of open repositories and Q\&A platforms~\citep{foalem2024studying, yahmed2023deploying}. Such studies may take various forms, including interviews, telephone surveys, or self-administered questionnaires. In this work, we opted for an \emph{online, self-administered questionnaire}, as this format enables access to geographically diverse participants, supports scalable data collection, and allows faster feedback from senior practitioners who often face time constraints. This approach has been widely adopted in prior empirical software engineering research~\citep{kitchenham2008personal, campbell2013coding}.
    
    \paragraph{Survey Design.}
    The survey was implemented using \emph{Google Forms\footnote{\url{https://docs.google.com/forms/u/0/}} }, a widely used platform for creating and distributing web-based questionnaires. The complete survey instrument is provided as supplementary material and summarized here for reproducibility~\citep{replication}. The questionnaire was administered in English to maximize participation from an international audience and is structured into nine sections, including an introductory consent page.
    
   The first section collects demographic and background information about participants, including their professional role, years of experience in software engineering, primary programming languages, and system architectures. This information allows us to characterize the participant pool and interpret responses in light of the professional context.
    
    Sections~2 to~7 focus on triangulating \textbf{RQ2} and \textbf{RQ3} by capturing practitioners' perceptions of the \emph{relevance} and \emph{difficulty} of individual logging topics identified in our Stack Overflow analysis. To reduce ambiguity, each logging topic was accompanied by a short description derived from the Stack Overflow analysis. Each topic is evaluated independently using Likert-scale questions to avoid aggregation bias and to enable direct comparison with Stack Overflow--derived indicators. These sections also include questions on the perceived engineering effort required to implement logging correctly. Participants are asked to compare logging effort against other common software engineering activities (e.g., refactoring, debugging, or CI/CD configuration). Mandatory open-ended questions invite respondents to justify their ratings and to describe technical, architectural, organizational, or tooling-related factors influencing logging difficulty.
    
    The final sections allow participants to report logging challenges not covered by the predefined topics and to share general reflections on what makes logging difficult--or manageable--in practice. Prior to dissemination, the survey instrument was reviewed internally by the authors to ensure clarity, completeness, and alignment with the research questions.

    Nevertheless, the survey instrument has inherent limitations. The perceived difficulty and effort associated with a logging topic may depend on contextual factors not fully captured by the questionnaire, such as specific frameworks, cloud providers, deployment environments, observability maturity, use of AI-assisted development tools, or project scale. Consequently, the survey findings should be interpreted as indicative of practitioner perceptions rather than universally generalizable or objective measures of logging complexity.
    
    \paragraph{Participant Recruitment.}
    To recruit practitioners with substantial industrial experience in software development, we relied on professional social networking platform, \emph{LinkedIn}, which is commonly used in empirical software engineering research to reach experienced industry practitioners \citep{yahmed2023deploying, foalem2024studying}. Our recruitment strategy combined public announcements, community-based dissemination, and targeted direct outreach.
    
    We first published a public call for participation on the first author's LinkedIn profile, which was subsequently shared by the other authors. The announcement was also disseminated within several LinkedIn groups dedicated to software engineering, DevOps, and machine learning. It explicitly invited senior and experienced practitioners and described the scope of the study, focusing on logging practices and challenges in modern software systems (e.g., containerized environments, CI/CD pipelines, testing infrastructures, and ML workflows). Participation was clearly described as voluntary, fully anonymous, and intended exclusively for academic research, with an estimated completion time of approximately 10 minutes. This initial outreach resulted in 5 responses collected over a two-week period.

    To further increase participation from practitioners with relevant senior-level experience, we conducted targeted direct outreach via LinkedIn private messages. Using LinkedIn's search functionality, we identified profiles matching keywords such as \emph{Senior Software Engineer}, \emph{Senior Developer}, \emph{Staff Engineer}, \emph{Tech Lead}, and \emph{Senior ML Engineer}. We then sent personalized, non-intrusive messages inviting these practitioners to participate in the survey. Each message briefly introduced the academic context of the study, clarified that the recipient was contacted based on publicly available professional information, and reiterated the voluntary and anonymous nature of participation. No follow-up messages were sent to individuals who did not respond. This targeted outreach yielded an additional 7 responses.
    
    \paragraph{Analysis of Survey Data.}
    We collected 12 responses from practitioners with substantial hands-on experience in logging-intensive software systems. All respondents met the minimum of 5 years of experience. The sample includes senior industrial roles (e.g., \emph{Senior Software Developer}, \emph{Machine Learning Engineer}, \emph{Tech Lead}, \emph{SRE} and \emph{Researcher \& Development}).  Regarding team organization, 41.7\% reported working in \emph{small teams} of 5 developers or fewer, while 50\% worked in \emph{medium-sized teams} comprising between 6 and 20 developers and 8.3\% worked in teams with \emph{more than 20 developers}.
    From a technical standpoint, respondents reported using a wide range of programming languages, most prominently Python and Java, along with C/C++ and JavaScript/TypeScript. Their professional experience spans monolithic architectures, microservices-based and distributed systems, as well as end-to-end machine learning pipelines covering training, inference, monitoring, and MLOps workflows. Consistent with this architectural diversity, participants reported familiarity with a broad set of logging and observability technologies, including traditional logging frameworks (e.g., Log4j, Logback, SLF4J, and Python's logging ecosystem), observability stacks (e.g., OpenTelemetry and ELK/Elastic Stack), cloud-native logging services, and machine learning--specific logging tools such as MLflow, TensorBoard, and Weights \& Biases.
\end{itemize}

\section{Results}
\label{sec:result}
This section reports about the results of our investigation into logging evolution on SO. The findings are structured around three key research questions, each aiming to shed light on different aspects of logging. These include the trends in logging discussions over time (RQ1), the primary topics within those discussions (RQ2), and the specific challenges faced by developers related to logging (RQ3). 

\subsection{RQ1: Trend of Logging}
\label{sec:RQ1}
In total, we extracted 216,094 logging-related posts (compared to 27,728 CI-related posts and 9,489 refactoring-related posts in 2019). Among the logging posts, 18,196 (54.7\%) received at least one answer, and 87,197 (40.4\%) had an accepted answer, contributed by 146,840 distinct developers. By contrast, refactoring-related questions had the highest percentage of accepted answers (∼64.41\%), while 8.73\% of them remained unanswered, involving 7,795 developers. For CI-related questions, 46.2\% had an accepted answer, while 36.7\% received at least one (but not accepted) answer, from 17,992 distinct developers \citep{peruma2022refactor, ouni2023empirical}.

\begin{figure}
    \centering
    \includegraphics[width=\textwidth]{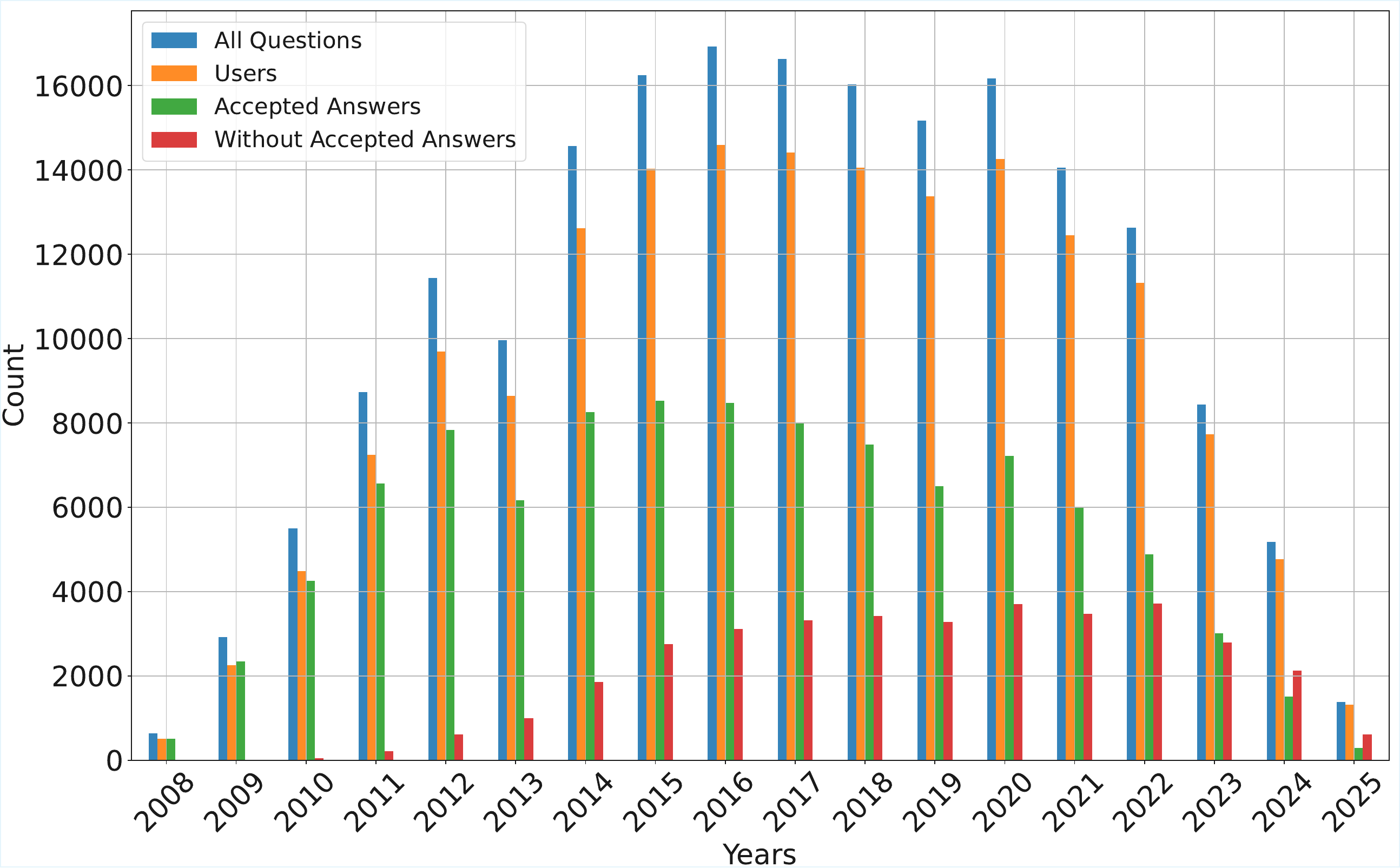}
    \caption{Popularity trend of logging post on SO over years.}
    \label{fig:popularity}
\end{figure}

\autoref{fig:popularity} illustrates the annual distribution and trends of logging-related posts and user activity on SO, covering the period 2008-2025. The data is categorized into four groups: (1) total questions, (2) unique users, (3) questions with accepted answers, and (4) questions without accepted answers. Below, we analyze the trends for questions and users.

\textit{Trends in Questions.}  
The blue bars show the overall number of logging questions posted annually. We observe a steady growth in question volume from 2008 until 2020, after which activity plateaued between 2021 and 2025. This reflects a strong and sustained interest in logging-related issues within the SO community, in line with previous findings up to 2019 \citep{peruma2022refactor, ouni2023empirical}.  

The green bars represent questions with accepted answers. From 2008 to 2012, the number of accepted answers increased notably. However, despite continued growth in question volume, the trend stabilized between 2013 and 2020, followed by a decline from 2021 onwards. A similar dynamic was reported for CI-related posts, which grew from 2009 to 2012, then stabilized until 2019 \citep{ouni2023empirical}. For refactoring-related posts, a brief two-year growth was observed from 2009 to 2011, followed by a sharp decline until 2019.  

The red bars show questions without accepted answers. Strikingly, from 2021 onward, the number of such questions approached that of questions with accepted answers, and by 2024 it exceeded them. This trend mirrors findings from refactoring-related studies, where unanswered questions eventually outnumbered those with accepted answers in the later years, with a near balance in the preceding three years. In contrast, CI-related posts experienced this imbalance much earlier, with unanswered questions outnumbering accepted ones roughly five years before 2019.  
The rise of unanswered questions suggests that community support is struggling to keep pace with newer or more complex challenges, which can be investigated by future study. Moreover, the broader similarity with trends in CI and refactoring indicates that this is not confined to logging, but reflects systemic shifts in SO usage and discussion dynamics. 

\textit{Trends in Users.}  
The orange bars indicate the number of unique users involved in logging discussions, either as askers or responders. User participation follows a trajectory similar to the total number of questions, with steady growth until 2020. From 2021 onward, both question volume and user engagement show a gradual decline. This downturn may be attributed to external factors, which could be investigated by future study, particularly the emergence of generative AI models that are increasingly capable of providing rapid and effective solutions to developer issues \citep{zhong2024can, da2024chatgpt}, or the migration of discussion to alternative platforms (e.g., GitHub Issues, Slack/Discord communities) highlight a redistribution of developer knowledge flows \citep{rong2018logging}.

\textit{Trends in Tags.}  
Our dataset contains a total of 20,758 distinct tags. Unsurprisingly, the most frequent tag is ``logging'', which appears in 48,965 questions (7.24\% of the dataset). To better understand the evolution of logging practices, we further examined the top 10 most frequent tags, excluding ``logging'' itself (Table~\ref{tab:popular_tags}). These tags fall into two broad groups: programming languages (Java 4.82\%, Python 2.52\%, C\# 2.18\%, JavaScript 1.72\%, PHP 1.33\%, Android 1.33\%) and logging frameworks (Log4j 2.75\%, Log4j2 1.30\%, Logstash 1.29\%, Logback 1.07\%). Together, these top 10 tags account for 20.31\% of all logging-related questions.  

\begin{figure}
    \centering
    \includegraphics[width=0.8\textwidth]{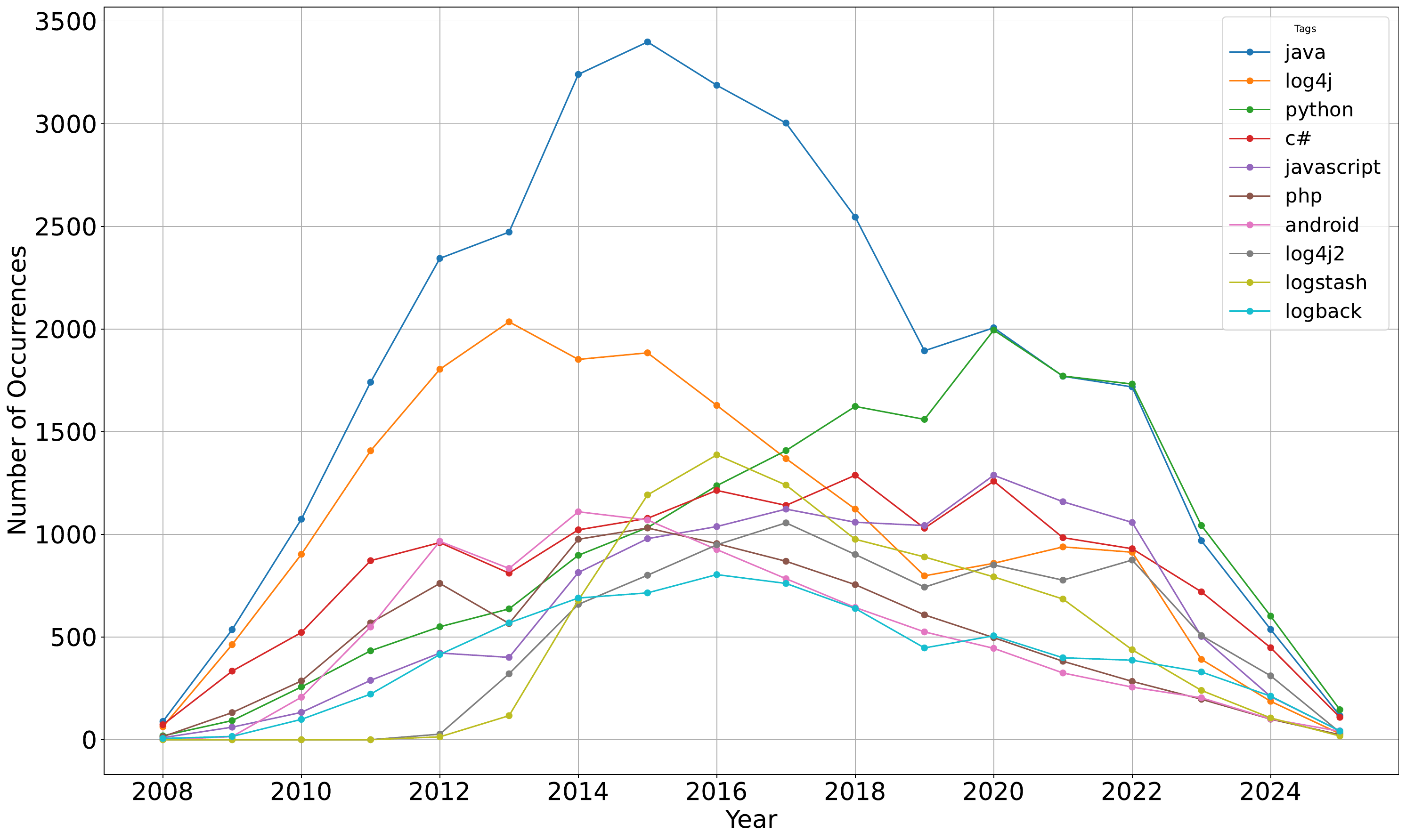}
    \caption{Logging tags popularity growth over years.}
    \label{fig:tags}
\end{figure}

\autoref{fig:tags} illustrates the temporal evolution of these tags on SO between 2008 and 2025. Several distinct patterns emerge. Java is the most dominant tag, peaking around 2015 before experiencing a fluctuating decline, reflecting both its longstanding prominence in enterprise programming language solutions and the gradual shift toward newer logging frameworks. Log4j exhibits two distinct waves: an early rise until 2013, followed by a dip and a second pronounced peak around 2020, after which it declined again. Python shows a more gradual increase, peaking between 2018 and 2020, coinciding with the broader adoption of Python in enterprise systems and data-intensive applications that rely heavily on logging, before tapering off \citep{gonzalez2020state}. C\# and JavaScript follow similar trajectories, with peaks between 2018 and 2021 before declining. PHP and Android peak earlier (2013-2015) and then steadily decline, consistent with the maturation and stabilization of those ecosystems. Among logging-specific frameworks, Log4j2 emerges around 2012, peaks between 2016 and 2017, and then declines. Logback and Logstash both rise after 2010, peak between 2015 and 2017, and subsequently decrease.  

Notably, nearly all series show sharp declines in the 2021-2025 window. By 2024-2025, most tags converge toward consistently low counts. This contraction parallels the broader decrease in logging-related questions, accepted answers, and user participation during the same period. The dominance of Java reflects its historical centrality in enterprise logging discussions, but its post-2016 decline was not directly offset by the growth of alternatives such as Log4j2, Logback, or Logstash. The earlier peaks for Android and PHP suggest that logging practices in these ecosystems reached a “steady state” sooner, whereas the later peaks for Python and C\# reflect their increasing importance in enterprise contexts during the late 2010s.  
  
These tag dynamics highlight how technological advances, ecosystem maturity, and shifts in programming language adoption shape logging practices. The variability in tag trends indicates that logging discussions on SO are not static but evolve with the rise and fall of programming languages, frameworks, and platforms. Our findings also align with broader adoption of programming language popularity \citep{programming}, suggesting that the growth and decline of logging tags are strongly correlated with the adoption curves of their associated technologies.
 
When contrasted with prior empirical studies on Continuous Integration (CI) and refactoring, we observe both similarities and differences. In CI-related discussions, the most popular tags center on CI servers and platforms (e.g., Jenkins, TeamCity, GitLab, Docker). These exhibit lifecycle dynamics: older tools (Hudson, Jenkins) decline while newer cloud-native tools (GitLab, Azure DevOps, Docker) rise \citep{ouni2023empirical}. Refactoring-related tags, by contrast, are dominated by programming languages (Java, C\#, JavaScript, Python, Ruby, PHP), with trends closely aligned to language programming popularity: Ruby and PHP peaked early and declined, while Python and JavaScript rose in the late 2010s \citep{peruma2022refactor}. Logging combines both dynamics: it is simultaneously shaped by the popularity of programming languages (Java, Python, C\#) and the adoption cycles or disruptions of logging frameworks (Log4j, Logback, Logstash). The ``double wave'' observed in Log4j, for example, is a unique phenomenon absent in CI or refactoring, reflecting how external shocks (e.g., critical vulnerabilities) can reignite discussions \citep{10628102}. 
 
To further investigate the technologies and concepts associated with logging discussions, we conducted a manual classification of tags. From the full set, we reviewed a statistically significant sample of 645 tags, which provides a 99\% confidence level with a 5\% confidence interval. Two authors independently suggested categories, drawing inspiration from prior classification studies \citep{ouni2023empirical, peruma2022refactor}. Categories were then reconciled and finalized through joint discussion, following the methodology of previous work \citep{ouni2023empirical, peruma2022refactor}.  

As a result, we identified ten clusters of tags: \textit{Programming Concepts, Framework/Library, Tool/IDE, Platform/Servers, Other, Database Concepts, Operating Systems (OS), Machine Learning Concepts, DevOps Tools, Programming Languages, and Security Concepts}. This categorization illustrates the diversity of technologies involved in logging discussions and highlights how logging evolves across different computing environments.

\begin{figure}
    \centering
    \includegraphics[width=0.8\textwidth]{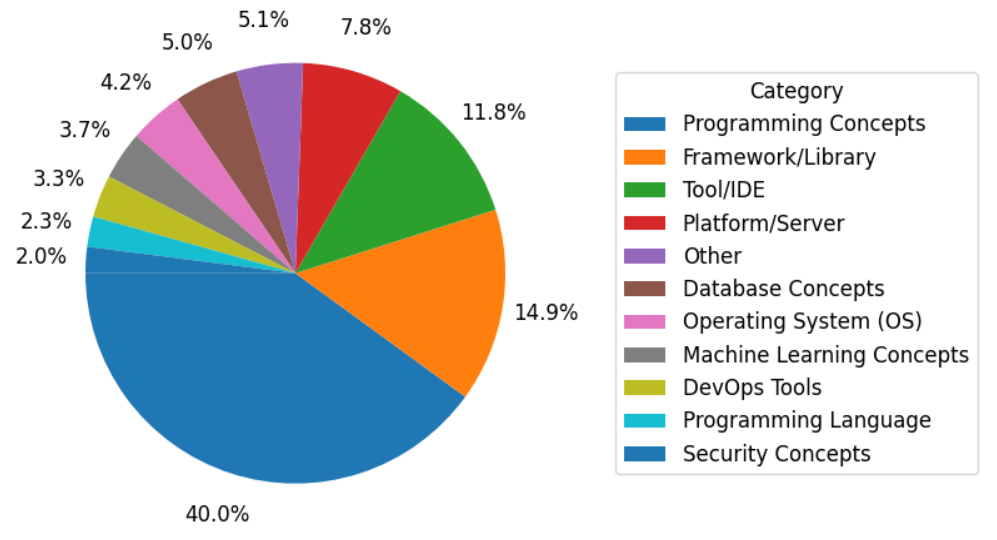}
    \caption{Distribution of concepts associated with logging.}
    \label{fig:tags_concepts}
\end{figure}

Figure \ref{fig:tags_concepts} shows the distribution of tags associated with logging. The majority of tags are related to programming concepts (40\%), such as ``exceptions" and ``biginteger", etc. Frameworks and libraries are the second most prevalent categories linked to logging (14.9\%), with examples including Symfony, Laravel, and Log4j. Tools and IDEs, such as Pentaho, VSCode, and Eclipse, account for 11.8\% of the tags, indicating their significant role in logging practices. The category Platform/Server (e.g., AWS, Tomcat, Apache, etc) comprises 7.8\% of the tags associated with logging. Database concepts represent 5\% of the tags, featuring items like ``submitchanges" and ``delta-live-tables". Operating systems such as Windows, iOS, and RedHat are linked to 4.2\% of the logging-related tags. Machine learning tools and concepts, including YOLO, Seq2Seq, and data science, contribute to 3.7\% of the tags. DevOps tools such as Git, TeamCity, Docker, and Ansible make up 3.3\% of the category. Popular programming languages like Python, Java, Android, and PHP fall under the Programming Language category, accounting for 2.3\% of the tags. The final category, Security Concepts, includes tags such as \textit{security, protected, and credential-manager}, which comprise 2\% of the distribution. This categorization highlights the diverse technologies and areas that integrate with logging practices across various software development environments.  \\
When compared to the distribution of CI and refactoring tags, several differences emerge. In the CI study, the dominant categories were \textit{platforms and servers} (40.15\%) and \textit{concepts} (20.52\%), with relatively fewer tags linked to programming languages (8.5\%) \citep{ouni2023empirical}. This reflects CI's strong association with server infrastructure (e.g., Jenkins, TeamCity, Docker) and process-oriented concepts (e.g., continuous deployment). In contrast, the refactoring study reported a predominance of \textit{algorithms and programming concepts} (43.22\%) and \textit{framework/library/API} tags (21.61\%), with programming languages (10.26\%) also playing a more substantial role \citep{peruma2022refactor}. Logging occupies a middle ground between these two domains. Like refactoring, it shows a heavy reliance on programming concepts. Unlike CI, where platform/server categories dominate, logging discussions are more evenly distributed across multiple categories, frameworks, IDEs, servers, databases, and even machine learning. This diversity underscores the fact that logging is not confined to infrastructure or language issues alone but cuts across the entire software stack, from low-level programming concepts to emerging fields such as ML and DevOps. \\  
The presence of machine learning and security-related tags further indicates that logging is increasingly tied to modern concerns such as AI monitoring \citep{bodor2023machine}, observability \citep{sakinala2025monitoring}, and trustworthy system design \citep{foalem2025logging} dimensions less visible in prior CI and refactoring studies. This diversity highlights the evolving role of logging as both a cross-cutting concern and a foundational practice for emerging domains.\\
To understand community engagement in logging discussions on Stack Overflow, we investigated the distribution of users participating in both questioning and answering. We can observe in Table \ref{tab:post_distribution}, that the majority of users (75.99\%) have posted only once, highlighting a trend where many engage primarily to seek answers to specific logging-related questions. This one-time posting behavior is accompanied by a sharp decline in users who post more frequently; only 16.26\% of users have posted twice, and fewer than 4\% have posted three times. This pattern is consistent, with the percentage of users decreasing as the number of posts increases, as presented in \autoref{fig:popularity}. Regarding answers, 76.81\% of users who have their answers accepted have only one accepted answer, suggesting that while many contribute valuable solutions, they do not regularly participate in logging discussions. This trend implies a potential lack of deep, sustained expertise and mentorship within the logging discussion community on SO. Similarly, the percentage of users with non-accepted answers follows a comparable trend, where 80.48\% of users do not receive acceptance for their single contribution. This community engagement observation isn't specific to logging and can also be observed in previous studies \citep{ouni2023empirical, peruma2022refactor}.

\begin{table}[htbp]
\centering
\caption{Top-10 most popular tags.}
\label{tab:popular_tags}
\begin{tabular}{@{}lrr@{}}
\toprule
Tag          & Occurrence & Percentage \\ \midrule
java         & 32,640     & 4.82\%     \\
log4j        & 18,647      & 2.75\%     \\
python       & 17,037      & 2.52\%     \\
c\#          & 14,795      & 2.18\%     \\
javascript       & 11,633      & 1.72\%     \\
php      & 9,007      & 1.33\%     \\
android        & 9,002      & 1.33\%     \\
log4j2         & 8,815      & 1.30\%     \\
logstash  & 8,775      & 1.29\%     \\ 
logback  & 7,259      & 1.07\%     \\ \bottomrule
\end{tabular}
\end{table}

\begin{table}[htbp]
\centering
\begin{threeparttable}
\caption{Distribution of the number of posts created by a user.\tnote{a}}
\label{tab:post_distribution}
\begin{tabular}{@{}lrr@{}}
\toprule
\textbf{Number of posts created by a user} & \textbf{Count} & \textbf{Percentage}\tnote{b} \\ 
\midrule
\multicolumn{3}{l}{\textit{Questions — total distinct users: 146{,}840\tnote{c}}} \\
1 & 111{,}584 & 75.99\% \\
2 & 23{,}871  & 16.26\% \\
3 & 5{,}108   & 3.48\% \\
4 & 2{,}775   & 1.89\% \\
5 & 1{,}186   & 0.81\% \\
Others\tnote{d} & 2{,}316 & 1.58\% \\ 
\midrule
\multicolumn{3}{l}{\textit{Accepted Answers — total distinct users: 68{,}652\tnote{e}}} \\
1 & 52{,}731 & 76.81\% \\
2 & 11{,}284 & 16.44\% \\
3 & 2{,}150  & 3.13\% \\
4 & 1{,}217  & 1.77\% \\
5 & 464      & 0.68\% \\
Others\tnote{d} & 806 & 1.17\% \\ 
\midrule
\multicolumn{3}{l}{\textit{Non-Accepted Answers — total distinct users: 89{,}835\tnote{f}}} \\
1 & 72{,}299 & 80.48\% \\
2 & 13{,}358 & 14.87\% \\
3 & 2{,}074  & 2.31\% \\
4 & 1{,}103  & 1.23\% \\
5 & 419      & 0.47\% \\
Others\tnote{d} & 582 & 0.65\% \\
\bottomrule
\end{tabular}
\begin{tablenotes}[flushleft]\footnotesize
\item[a] The table reports distributions over \emph{distinct users} per role (askers, users with accepted answers, users with non-accepted answers), not counts of posts. 
\item[b] Percentages are computed \emph{within each block} (Questions / Accepted Answers / Non-Accepted Answers) using that block's “total distinct users” as the denominator.
\item[c] “Questions” counts each distinct user who asked at least one logging-related question and groups them by how many questions they asked.
\item[d] “Others” aggregates users with $\ge 6$ contributions in the corresponding role.
\item[e] “Accepted Answers” counts each distinct user who has at least one answer marked as accepted and groups them by how many \emph{accepted} answers they authored.
\item[f] “Non-Accepted Answers” counts each distinct user who authored answers that were \emph{not accepted} and groups them by how many non-accepted answers they authored.
\end{tablenotes}
\end{threeparttable}
\end{table}
\begin{tcolorbox}[colback=black!10!white,colframe=black]
\textbf{Findings:} 
Our analysis reveals a proportional trend between logging questions and user engagement on Stack Overflow, with many contributors providing accepted answers, though most participate only once. A similar engagement pattern is observed for Continuous Integration and refactoring on Stack Overflow. Distinctively, logging emerges as a cross-cutting topic, spanning programming languages, frameworks, DevOps tools, machine learning, and security. This hybrid nature sets it apart from Continuous Integration, which is infrastructure-driven, and refactoring, which is programming language and design driven, underscoring logging's widespread and evolving relevance across diverse technological domains.  
\end{tcolorbox}

\subsection{RQ2: Popular Topic of Logging}
\label{sec:RQ2}
\begin{figure}
    \centering
    \includegraphics[width=0.6\textwidth]{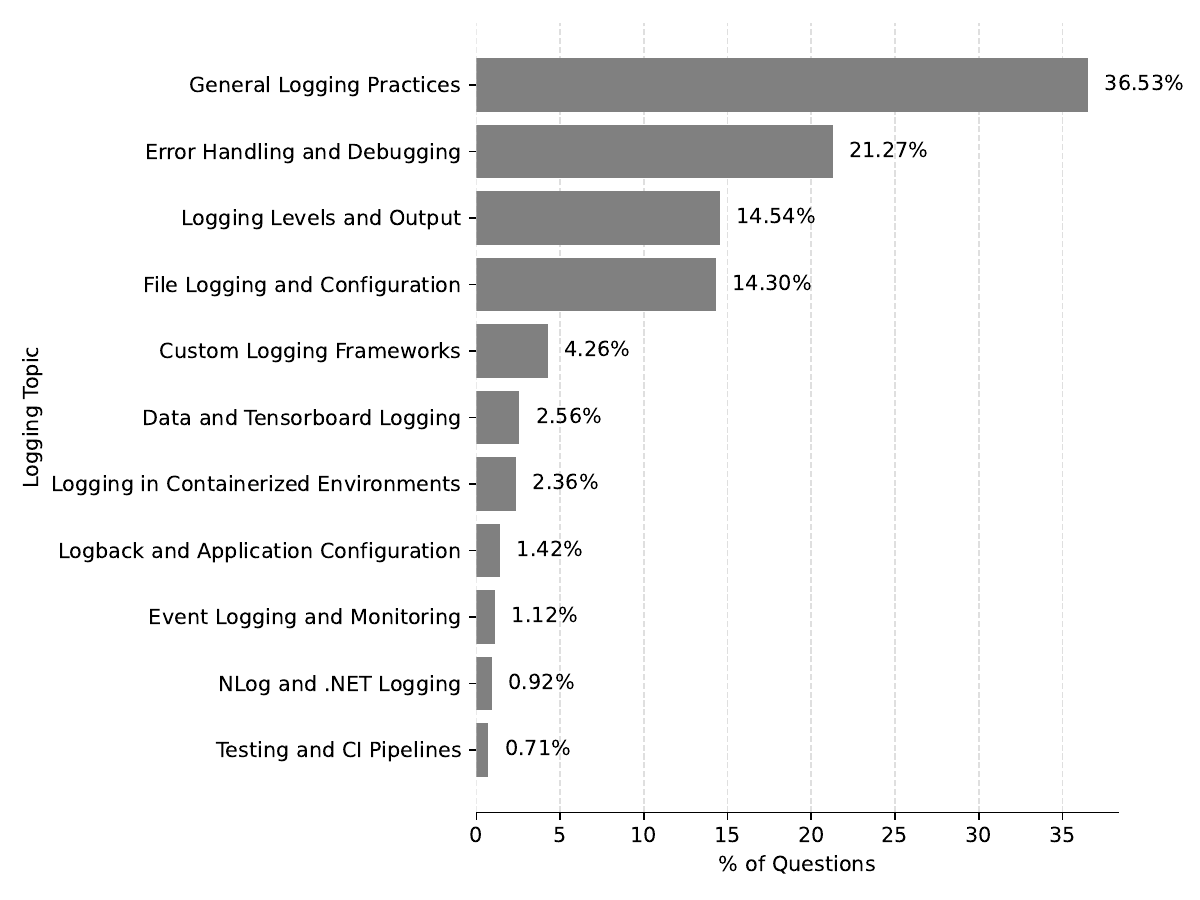}
    \caption{Logging topics by the percentage of their corresponding questions.}
    \label{fig:horizontal_topic}
\end{figure}

We utilized an LLM-based approach to identify prevalent topics within Stack Overflow logging discussions. This method allowed us to classify discussions into well-defined topics based on both the title and body of the posts. As shown in \autoref{fig:horizontal_topic}, the analysis revealed 11 distinct topics. \textit{General Logging Practices} emerged as the most discussed topic, comprising \textbf{36.53\%} of questions (78{,}939 of 216{,}094). It is followed by \textit{Error Handling and Debugging} at \textbf{21.27\%} (45{,}963) and \textit{Logging Levels and Output} at \textbf{14.54\%} (31{,}420). A further \textbf{14.30\%} (30{,}901) of posts concern \textit{File Logging and Configuration}. The remaining topics each account for less than 5\% of questions: \textit{Custom Logging Frameworks} (4.26\%), \textit{Data and TensorBoard Logging} (2.56\%), \textit{Logging in Containerized Environments} (2.36\%), \textit{Logback and Application Configuration} (1.42\%), \textit{Event Logging and Monitoring} (1.12\%), \textit{NLog and .NET Logging} (0.92\%), and \textit{Testing and CI Pipelines} (0.71\%). 
These topics are consistent with, and extend, the findings of \citet{gujral2021exploratory}, who conducted an exploratory semantic analysis of logging-related Stack Overflow posts using LDA-based topic modeling. Their study identified dominant logging concerns related to programming languages (e.g., Java and Python), object-oriented programming constructs (e.g., classes, methods, and inheritance), and general logging usage patterns. Our results confirm the persistence of these core concerns captured here by topics such as \textit{General Logging Practices}, \textit{Error Handling and Debugging}, and \textit{Logging Levels and Output,} and reveal newer, ecosystem-specific areas (e.g., containerized environments, CI pipelines, and ML logging).
These topic distributions can also be understood in light of RQ1 (Section \ref{sec:RQ1}). As seen in \autoref{fig:popularity}, logging discussions grew steadily until 2020 before plateauing and later declining, a shift mirrored in the declining adoption of dominant tags such as Java and Log4j. This temporal contraction helps explain why only a small subset of topics, such as \textit{General Logging Practices}, \textit{Error Handling and Debugging}, and \textit{Logging Levels and Output}, continued to dominate discussions, because they (i) recur across stacks and skill levels, (ii) are tightly coupled to day-to-day development (configuration, failure diagnosis, message/level choices), while specialized areas like \textit{Data and TensorBoard Logging} and \textit{Event Logging and Monitoring} remained relatively marginal, because they (a) involve narrower practitioner populations (ML practitioners) and (b) depend on tool-specific ecosystems (TensorBoard,MLflow). Moreover, the tag analysis in RQ1 \autoref{fig:tags} showed that programming concepts (40\%) and frameworks/libraries (14.9\%) dominate the ecosystem of logging-related discussions. These broad categories strongly overlap with the most prevalent topics identified.
In the following, we detail each topic and provide representative questions.

\textit{\textbf{1. General Logging Practices (36.53\%).}} This is the most discussed topic (\autoref{fig:horizontal_topic}), includes questions related to the general usage of logging across various programming languages such as Python (e.g., $Q_{\text{id}}$: 78901762) \citep{stack5}, JAVA (e.g., $Q_{\text{id}}$: 4587174) \citep{stack6}, .NET (e.g., $Q_{\text{id}}$: 78860915) \citep{stack7}, and Android (e.g., $Q_{\text{id}}$: 5190860) \citep{stack8}. Previous studies have characterized logging practices in popular programming languages including Java \citep{chen2017characterizing}, Android \citep{zeng2019studying}, C/C++ \citep{yuan2012characterizing}, identifying pervasive trends and common challenges. As a significant area of interest in logging-related discussions, there is a clear need for more educational resources on the usage and implementation of logging across different facets of programming. \\ 
\textit{\textbf{2. Error Handling and Debugging (21.27\%).}} This topic, highlighted as the second most popular in \autoref{fig:horizontal_topic}, includes challenges and errors faced by practitioners when implementing logging in their projects. Questions such as (e.g., $Q_{\text{id}}$: 78819869) \textit{``Logging issue in my application after migrating from Wildfly 18 to 33"} and (e.g., $Q_{\text{id}}$: 78965763) \textit{``logging/print statements not being captured when running script with Runas command"} exemplify the difficulties encountered. Similar to previous studies \citep{li2020qualitative}, other questions in this category draw connections between logging and debugging, such as (e.g., $Q_{\text{id}}$: 78901762) \textit{``Visual Studio Console Debug Popup"} and (e.g., $Q_{\text{id}}$: 78901762) \textit{``Intellij IDEA getting remote logs while debugging"}. These inquiries highlight practitioners' awareness of the critical role that logging plays in the debugging process.\\ 
\textit{\textbf{3. Logging Levels and Output (14.54\%).}} This topic focuses on choosing log levels (DEBUG/INFO/WARN/ERROR) and composing messages/dynamic fields. Previous research indicates that setting the appropriate log level, message, and dynamic variables presents significant challenges for logging practitioners \citep{li2017log,anu2019approach,li2020towards}, which explains why practitioners often turn to Stack Overflow for assistance. Typical queries include issues related to log levels (e.g., $Q_{\text{id}}$: 2031163; 71354500) \citep{stack1,stack2}, log messages (e.g., $Q_{\text{id}}$: 14058453) \citep{stack3}, and dynamic variables (e.g., $Q_{\text{id}}$: 78958199) \citep{stack4}.\\
\textit{\textbf{4. File Logging and Configuration (14.30\%).}} This topic focuses on the technical aspects of managing and configuring log files effectively. It includes common questions related to the organization and storage of log data, such as (e.g., $Q_{\text{id}}$: 78743357) \textit{``How to Archive log4net Logs to an Archive Folder with Date and Time in Filename?"} which involves time-stamped log archiving. Other inquiries like (e.g., $Q_{\text{id}}$: 78972736) \textit{``Splitting up the Log Files in Log4j2?"} address the partitioning of logs for better manageability and performance. Additionally, questions such as (e.g., $Q_{\text{id}}$: 78979109) \textit{``Logs not being written if the parent directory is included in the filename"} reflect challenges in log file path configurations.\\
\textit{\textbf{5. Custom Logging Frameworks (4.26\%).}} This topic explores how practitioners often require or develop custom logging solutions in response to the unique needs of their applications. For example, developers seek guidance on integrating custom functionality within established logging libraries, as evidenced by questions like (e.g., $Q_{\text{id}}$: 78863938) \textit{``Is it possible to access the logger name in a custom logproc of the Tcl logger library?"} and (e.g., $Q_{\text{id}}$: 78753528) \textit{``How to use a custom DB logger with TypeORM?"}. These queries highlight the need for customization in logging practices, allowing for enhanced functionality such as accessing logger attributes or integrating logging with database management systems. Furthermore, even within widely used languages like Python, there is a demand for adapting traditional logging approaches to suit specific project requirements. An illustrative question from the SO community is (e.g., $Q_{\text{id}}$: 78613726) \textit{``Python custom logger multiple modules with variable logger name"} \citep{stack9}, where the challenge lies in extending the Python logging library to handle variable logger names across multiple modules. This topic underscores the ongoing need for flexible logging solutions that can adapt to diverse development environments and requirements.\\
\textit{\textbf{6. Data and TensorBoard Logging (2.56\%).}} Here, discussions are related to logging within data-intensive systems, particularly those involving machine learning applications. Practitioners explore tools like TensorBoard \citep{TensorBoard} and MLflow \citep{MLflow} to effectively manage and visualize logs in such environments. Questions such as (e.g., $Q_{\text{id}}$: 70478173) \textit{``How to track the big data stored in Gdrive through DVC?"}, (e.g., $Q_{\text{id}}$: 78753528) \textit{``Why does tensorboard not show all metrics?"}, (e.g., $Q_{\text{id}}$: 72684326) \textit{``How to change Sklearn flavors version in mlflow on azure machine learning?"}, underscore the complexity and the specific needs of logging in ML-based applications. The popularity of TensorBoard as a logging tool highlights its significance in providing insights that are crucial for the optimization and tuning of machine learning models. Representing 5.21\% of logging discussion on SO, there is a need for further research to investigate challenges faced by ML practitioners in logging practice. Recent studies, begin to address these gaps by investigating logging practices in ML-based applications \citep{foalem2024studying}.\\
\textit{\textbf{7. Logging in Containerized Environments (2.36\%).}} As depicted in Figure \ref{fig:tags_concepts}, logging has been associated with a variety of technologies and concepts, including concepts around containerized environments. Containerization technologies such as Docker and Kubernetes have revolutionized deployment practices, necessitating sophisticated logging mechanisms as highlighted by many previous studies \citep{alves2021identifying, chen2020docker}. Questions such as (e.g., $Q_{\text{id}}$: 78934939) \textit{``Filebeat: Not able to scrape logs of Spark on Kubernetes"} and (e.g., $Q_{\text{id}}$: 75725865) \textit{``How to Prevent Docker Container Logs from Disappearing?"} reflect common concerns about capturing and preserving logs in these environments. These discussions highlight the need for robust logging solutions that can adapt to the complexities of container orchestration, ensuring that logs are accessible and meaningful for debugging and monitoring distributed applications. \\
\textit{\textbf{8. Logback and Application Configuration (1.42\%).}} Logback, is a Java logging framework, that often requires extensive configuration to effectively manage log outputs, particularly when integrated with frameworks like Spring Boot. which frequently prompts developers to seek assistance from SO community. Common challenges include issues like (e.g., $Q_{\text{id}}$: 78441955) \textit{``Logback Spring Boot configuration is not working on Azure"}. Developers also encounter problems with basic Logback setups in Java applications, as seen in posts like (e.g., $Q_{\text{id}}$: 54467849) \textit{``Java based Logback Configuration"}\citep{stack10}. Moreover, the need for dynamic logging configurations leads to questions such as (e.g., $Q_{\text{id}}$: 77171429) ``How do I programmatically tell Logback 1.4.x to Reload Configuration?''. This ongoing dialogue within the SO community illustrates the indispensable role of peer support in overcoming the intricacies of Logback configuration.\\
\textit{\textbf{9. Event Logging and Monitoring (1.12\%).}} Event logging plays a pivotal role in monitoring applications by capturing and recording key events that occur within software systems. These logs are crucial for diagnosing issues, understanding software behavior, and improving system reliability. Prior research \citep{cinque2012event,vaarandi2008mining,
liang2007failure}, has highlighted the significance of event logs in analyzing software failures and enhancing fault diagnosis processes. In practical terms, developers are often confronted with specific challenges related to event logging, such as ensuring the synchronization of clock settings in distributed environments, as questioned in \textit{``Do AWS CloudWatch Logs Events have a Synchronized Clock?"} (e.g., $Q_{\text{id}}$: 52488775). Additionally, developers encounter issues with how event information is structured and recorded, exemplified by \textit{``OpCode is being attached to event name in out-of-process Semantic Logging (SALB) C\#"} (e.g., $Q_{\text{id}}$: 41671072). These inquiries illustrate the ongoing need to refine event logging techniques to support effective monitoring and fault resolution strategies in complex software environments.\\
\textit{\textbf{10. NLog and .NET Logging (0.92\%).}} This topic encompasses specialized discussions focused on leveraging the NLog library within .NET applications, addressing issues specific to this popular logging framework. For instance, developers encounter challenges such as configuring NLog to work effectively in modern frameworks, exemplified by the query (e.g., $Q_{\text{id}}$: 63065604) \textit{``NLog target database not working in ASP.NET Core 2.2 MVC"}.\\
\textit{\textbf{11. Testing and CI Pipelines (0.71\%).}} The integration of logging within testing and continuous integration (CI) pipelines is a critical intersection of development practices, aimed at enhancing diagnostic capabilities and ensuring robust software delivery. Although this is the least-discussed topic in our corpus, prior research has begun to tackle logging within test code and testing workflows (e.g., \citep{zhang2022studying}). This topic explores questions related to capturing and analyzing logs generated during automated testing and CI processes. For instance, developers inquire about issues such as \textit{``Where are the Selenium Logs when running in Azure DevOps Server"} (e.g., $Q_{\text{id}}$: 58218200) and how to \textit{``Customize Selenium logs for failed tests"} (e.g., $Q_{\text{id}}$: 39725477), which pertains to automated testing using the Selenium framework. Additionally, there is interest in real-time logging during test execution, as evidenced by queries like \textit{``jest, logging success or failure between each test while the tests are running"} (e.g., $Q_{\text{id}}$: 48417643). These discussions underscore the evolving role of logging in providing transparency and accountability in automated testing environments, which are crucial for troubleshooting and refining the CI/CD pipelines.\\
\subsubsection{Survey Validation of Topic Relevance}
\begin{figure}
    \centering
    \includegraphics[width=0.9\textwidth]{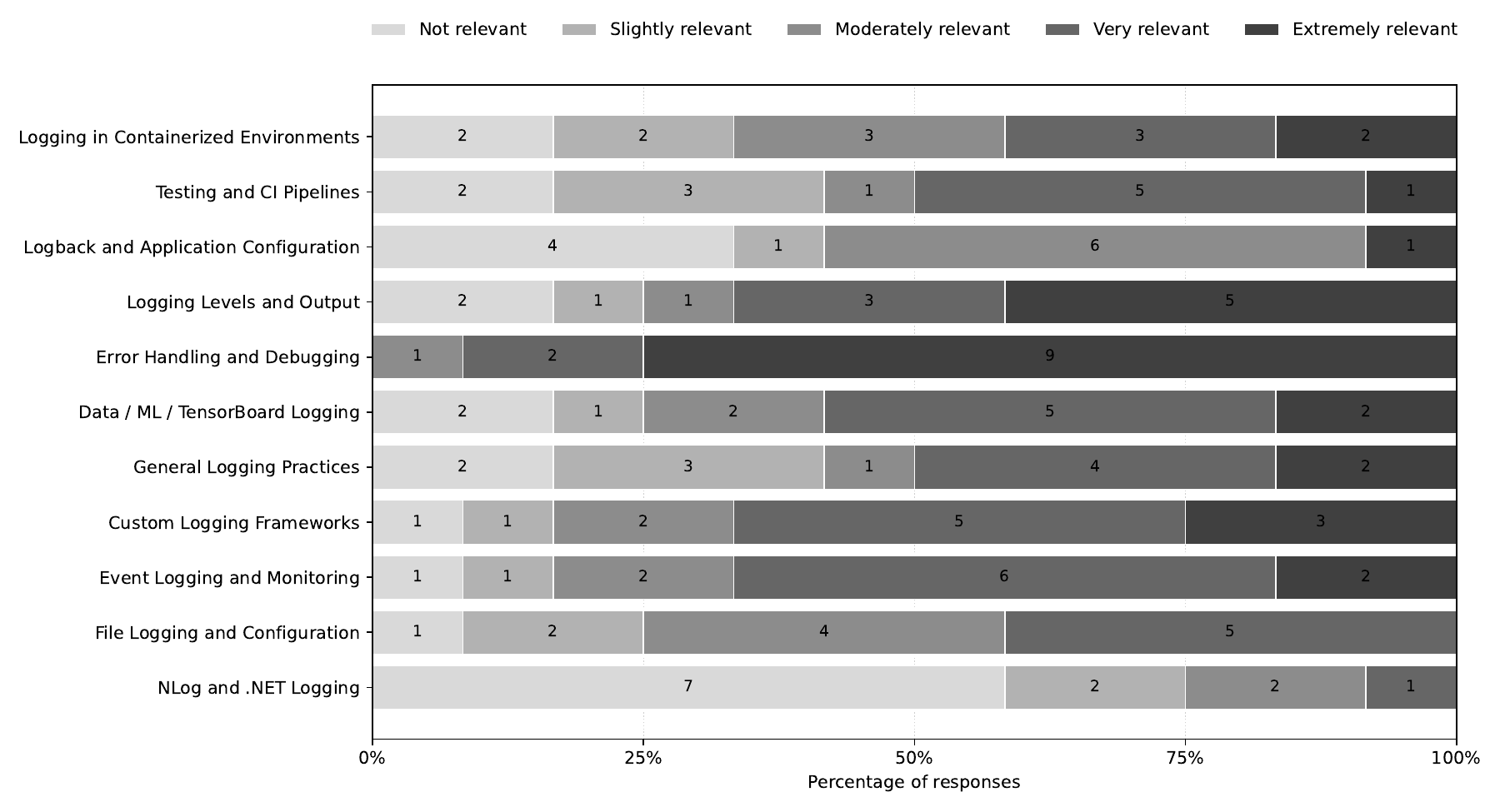}
    \caption{Distribution of survey responses on the industrial relevance of the logging topic.}
    \label{fig:survey_relevance}
\end{figure}

To validate whether the logging topics identified from Stack Overflow discussions (RQ2) align with industrial practice, we asked survey participants to assess the \emph{industrial relevance} of each topic using a five-point Likert scale ranging from \emph{Not relevant} to \emph{Extremely relevant}. In this study, \textit{relevance is defined as the extent to which a logging topic frequently and critically manifests in day-to-day industrial work}, that is, how regularly it affects system development, operation, debugging, or maintenance activities in practitioners' projects. Under this definition, a topic is considered highly relevant if it recurrently influences engineering decisions or imposes non-trivial effort during the software lifecycle. Figure~\ref{fig:survey_relevance} summarizes the distribution of responses across the 11 topics.

Overall, the results show a strong correspondence between topic popularity on Stack Overflow and perceived industrial relevance as defined above. Core logging topics--namely \textit{Error Handling and Debugging}, \textit{Logging Levels and Output}, and \textit{General Logging Practices}--were unanimously rated as at least \emph{Moderately relevant}, with the majority of respondents assigning \emph{Very} or \emph{Extremely relevant} ratings. In particular, \textit{Error Handling and Debugging} received the strongest endorsement, with 9 respondents ($75\%$) rating it as \emph{Extremely relevant}. This reflects its pervasive and critical role in diagnosing failures, understanding runtime behavior, and maintaining system reliability in production environments.\\
Topics related to configuration and operational concerns--such as \textit{Logback and Application Configuration}, \textit{File Logging and Configuration}, and \textit{Event Logging and Monitoring}--were also consistently perceived as relevant. For instance, \textit{Logback and Application Configuration} was rated as \emph{Moderately relevant} by 6 participants ($50\%$). Interpreted through our relevance definition, this suggests that while such topics may not arise continuously during routine development, they represent recurring and often time-consuming challenges when systems evolve, are deployed to new environments, or require reconfiguration.

More specialized topics exhibit greater variance in relevance assessments. Notably, \textit{Logging in Containerized Environments} and \textit{Testing and CI Pipelines}--which were among the least frequent topics in the Stack Overflow corpus--were nevertheless rated as \emph{Very relevant} or higher by a majority of respondents. This indicates that although these topics generate fewer public questions, they frequently and critically affect industrial workflows when they do arise, particularly in modern cloud-native and DevOps-oriented development settings.\\
In contrast, \textit{NLog and .NET Logging} stands out as the only topic predominantly rated as \emph{Not relevant} (7 respondents, $58.33\%$). This outcome aligns with the technological backgrounds of the surveyed practitioners, who primarily reported experience in Java-, Python-, and ML-centric ecosystems. Under our definition of relevance, this topic rarely affects their day-to-day work, which explains both its low industrial relevance ratings and its marginal presence in the Stack Overflow dataset.\\
Finally, \textit{Data/TensorBoard Logging}, while representing a small fraction of Stack Overflow discussions, was rated as at least \emph{Very relevant} by 7 respondents ($58.33\%$). This result reinforces the observation that ML-related logging issues, although less visible in general-purpose Q\&A platforms, frequently and critically impact practitioners working in data-intensive and MLOps settings, particularly during experimentation, monitoring, and model debugging.\\
Taken together, these results provide qualitative validation for the topic taxonomy derived in RQ2. When interpreted through a practitioner-grounded definition of relevance, the survey confirms that the most frequently discussed topics on Stack Overflow correspond to foundational industrial concerns, while also revealing that several low-frequency topics--especially those related to containerization, CI pipelines, and ML logging--carry disproportionate importance in real-world engineering practice.

\begin{tcolorbox}[colback=black!10!white,colframe=black]
\textbf{Findings:}
We identified 11 topics related to logging. \textit{``General Logging Practices''} emerged as the most discussed topic, while \textit{``Testing and CI Pipelines''} was the least discussed. The top three topics (\textit{``General Logging Practices''}, \textit{``Error Handling and Debugging''}, \textit{``Logging Levels and Output''}) collectively account for 72.34\% of the entire discussion on logging; adding \textit{``File Logging and Configuration''} brings the top four to 86.64\%. The practitioner survey  validates the industrial relevance of this topic taxonomy, confirming that the most prevalent Stack Overflow topics correspond to foundational logging concerns in practice. Moreover, the survey highlights that several low-frequency topics—particularly those related to containerized environments, CI pipelines, and ML logging—are nevertheless perceived as highly relevant by practitioners.
\end{tcolorbox}

\subsection{RQ3: Logging Challenges}
\label{sec:RQ3}
\autoref{tab:popular_difficulty} presents the results on topic popularity and difficulty. Popularity is quantified through two metrics: Average Views (Avg. Views) and Average Scores (Avg. Scores), whereas difficulty is assessed by the percentage of questions without accepted answers (No Acc. Answers \%), the percentage of unanswered questions (No Answers \%), and the median time to receive an accepted answer (Median Hours to Acc. Answer). These metrics and approaches are based on the methodologies presented in previous studies \citep{openja2020analysis, peruma2022refactor, ouni2023empirical}.

\begin{figure}
    \centering
    \includegraphics[width=\textwidth]{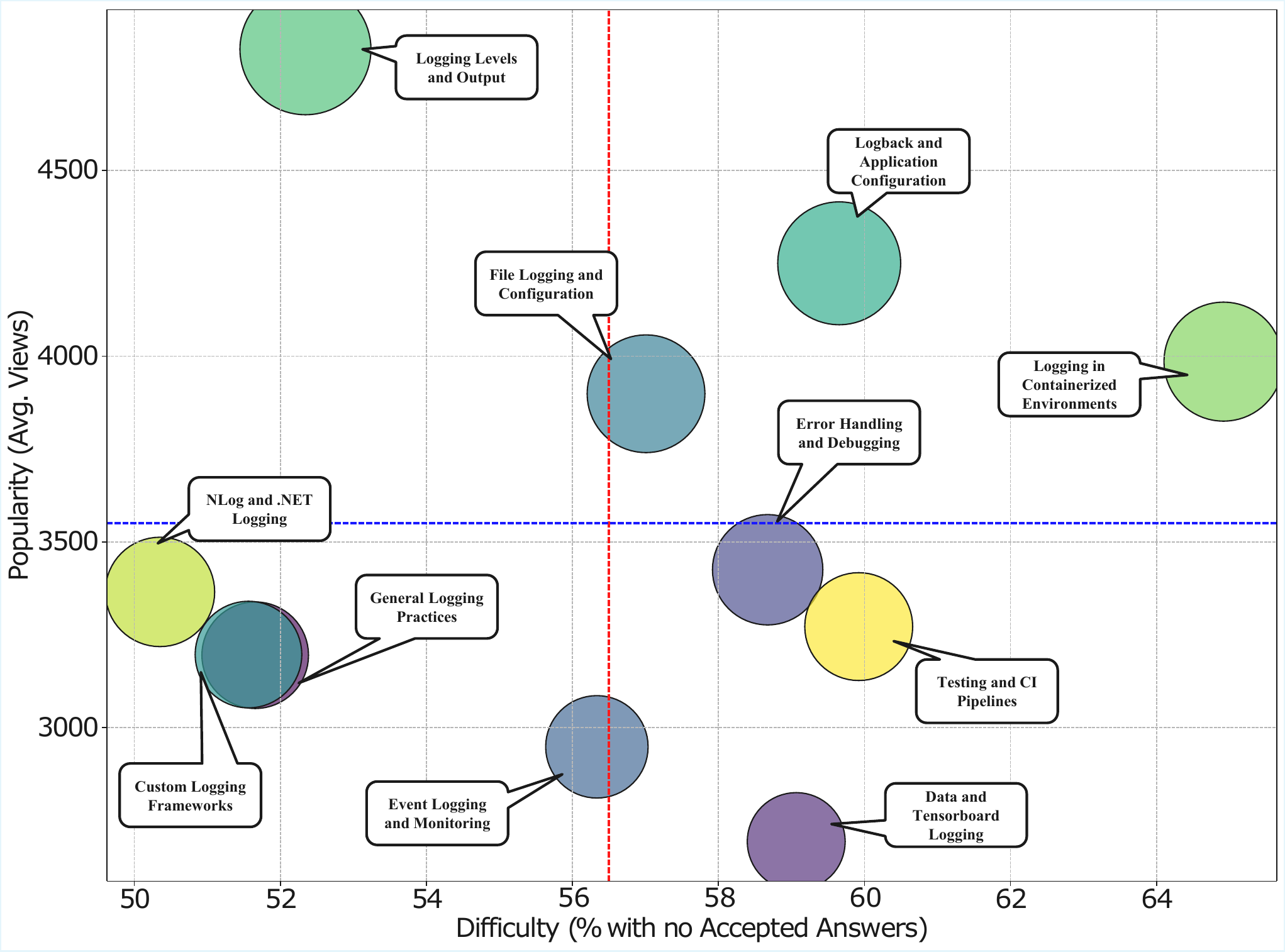}
    \caption{Comparison of logging topics by popularity \& difficulty}
    \label{fig:bubble}
\end{figure}

\textbf{Popularity:} As observed in \autoref{tab:popular_difficulty}, the topic ``Logging Levels and Output'' emerges as the most popular, with the highest average score of 4.13 and average views of 4825.73. In contrast, ``Data and Tensorboard Logging'' ranks as the least popular, closely followed by ``Event Logging and Monitoring''. ``Logback and Application Configuration'' and ``File Logging and Configuration'' also show high popularity, exhibiting significant viewership as depicted in Figure \ref{fig:bubble}.
These trends align with findings from prior studies on logging practices. For instance, researchers have noted that developers often struggle to determine appropriate log verbosity levels, making configuration of log levels a frequent topic of discussion \citep{li2017log, anu2019approach, kim2020automatic, li2020qualitative}. Likewise, the prominence of ``Logback and Application Configuration'' corresponds to the widespread adoption of the Logback framework in industry in 2024; over half of Java applications used Logback, which naturally leads many practitioners to seek guidance on configuring this tool \citep{NewRelic}.
``Data and TensorBoard Logging'' appears to be the least popular topic, suggesting that logging in data-driven systems is a niche concern among practitioners. Empirical studies have found that logging is less pervasive in machine learning-based applications than in traditional software \citep{foalem2024studying}. Meanwhile, the relatively lower interest in ``Event Logging and Monitoring'' may indicate that advanced log monitoring is a more specialized challenge, echoing observations that deriving insights from logs at scale remains difficult in practice \citep{candido2021log}.
In the CI study \citep{ouni2023empirical}, the most popular topics were ``Build'' and ``Culture'', both of which attracted the highest attention, with average scores of 3.28-3.23 and views 2488.8-2082.8, respectively. These results reflect CI's strong association with infrastructure and platform-oriented practices.
In the Refactoring study \citep{peruma2022refactor}, popularity was concentrated on ``Architecture and Design Patterns'' and ``Unit Testing'', which reached the highest average views (2383.3 and 2188.7) and scores (1.25-1.26). This pattern aligns with refactoring's language and design-centric nature, emphasizing code optimization, testing, and design best practices.\\
When comparing average popularity across the three domains, logging emerges as the most popular overall, with an average of 3550.56 views per topic and an average score of 2.67. This surpasses both refactoring, which averages 2747.3 views and a score of 1.25, and CI, which averages only 1996.5 views and a score of 2.94. These indicate that logging discussions attract broader developer attention than either CI or refactoring.

\textbf{Difficulty:} Regarding question difficulty, as shown in \autoref{tab:popular_difficulty} and illustrated in Figure~\ref{fig:bubble}, ``Logging in Containerized Environments'' is the most difficult topic: it has the highest share of questions without accepted answers (64.91\%) and the largest unanswered proportion (22.96\%), with a relatively long median time to acceptance (47.32 hours). The second most difficult topic by the same criterion is ``Testing and CI Pipelines'', with 59.92\% of questions lacking an accepted answer and a median of 48.54 hours. 
These difficulty patterns are consistent with prior large-scale analyses of developer Q\&A on Stack Overflow. For example, \citet{openja2020analysis} show that topics related to \textit{Docker}, \textit{continuous deployment}, and \textit{software testing} are among the hardest to resolve, suggesting that container-, testing-, and pipeline-centric discussions systematically attract more complex questions and receive weaker community support. Our findings extend this line of work by demonstrating that logging-related issues become particularly difficult when they arise within such environments.
Domain-specific studies help explain the sources of this difficulty. \citet{alves2021identifying} report that, in containerized Python applications, logging is tightly coupled with observability concerns such as distributed execution, log dispersion across services, and limited runtime visibility, all of which complicate debugging and diagnosis. Likewise, \citet{zhang2022studying} observe that logging in test code is often ad hoc and strongly tied to specific testing frameworks, increasing the effort required to diagnose failures in testing and CI settings.\\
At a technical level, logging is inherently cross-cutting, stateful, and environment-dependent. Unlike functional code, logging behavior arises from the interaction between application logic, runtime configuration, execution context, and external infrastructure. In containerized systems, this complexity is amplified by container ephemerality, distributed execution, and indirect log routing through sidecars, collectors, or orchestration layers. As a result, developers must reason not only about \textit{what to log}, but also \textit{where} logs are routed, \textit{when} they are persisted, and how they are correlated across services--questions that often lack a single, definitive answer and depend heavily on deployment architecture.
Similarly, in testing and CI pipelines, logging must serve multiple competing purposes simultaneously: diagnosing transient test failures, supporting automated execution, and remaining lightweight enough not to slow pipelines. \citet{zhang2022studying} show that logging in test code is frequently ad hoc, tightly coupled to specific frameworks, and poorly aligned with CI execution models, which obscures failure causes when tests run in parallel or in headless environments. Consequently, many logging-related questions in CI contexts remain unresolved, as the “correct” solution depends on subtle interactions among test frameworks, execution order, log levels, and CI infrastructure rather than on logging APIs alone.\\
``Logback and Application Configuration'' ranks third on this difficulty metric (59.65\% without accepted answers) and is distinctive for having by far the longest median time to an accepted answer (118.14 hours), pointing to persistent configuration challenges despite Logback's wide use in the JAVA programming language \citep{NewRelic}. 
At the other end of the spectrum, ``NLog and .NET Logging" appears the least difficult, with the lowest rate of questions without accepted answers (50.35\%) and a relatively long median resolution time (74.89 hours). ``General Logging Practices'' is also among the least difficult, showing a near-lowest rate (51.56\%) and the shortest median resolution time (21.36 hours). ``Custom Logging Frameworks'' follows closely with 51.65\% of questions lacking accepted answers. By contrast, ``Data and Tensorboard Logging'' though among the least popular, shows 59.06\% without accepted answers and a median of 22.00 hours, indicating that while questions are fewer, they are typically resolved relatively quickly when they arise.\\
``Logging Levels and Output'' also stands out as both the most popular topic and one of the least difficult ones. It attracts the highest attention from developers in terms of average score (4.13) and views (4825.73), yet only 52.37\% of its questions remain without accepted answers, and the fewest unanswered questions (15.49). This dual status reflects the fact that, while developers frequently discuss when to use levels such as DEBUG, INFO, WARN, or ERROR, the community is often able to provide guidance. The relative ease of resolving such questions is consistent with prior academic work: early research adopted machine learning to predict appropriate logging levels automatically, framing log severity assignment as a classification problem and demonstrating that patterns of log-level usage can be effectively learned from code repositories \citep{li2017log,zhu2015learning}.
When compared with related domains. In the CI study \citep{ouni2023empirical}, the most difficult topics were ``Version Control'' (59\% without accepted answers) and ``Deployment'' (58\%), while ``Testing'' had the longest median time to an accepted answer (18 hours).
In the refactoring study \citep{peruma2022refactor}, the most difficult topic were ``Tools and IDEs'' (42.34\% without accepted answers) and ``Unit Testing'' (37.78\%), while ``Database'' showed the longest median time to receive an accepted answer (0.44 hours). \\
Taken together, these comparisons reveal that logging-related discussions emerge as the most difficult and time-consuming, with an average of 56.49\% of questions lacking accepted answers and a median of 45.37 hours to reach resolution. CI questions are similarly difficult in terms of unresolved proportion (54\%), but they are typically resolved much faster, with a median time of only 11 hours. In contrast, ``Refactoring'' appears considerably less difficult overall, with just 36.58\% of questions lacking accepted answers and an extremely short median resolution time of 0.39 hours.

\begin{table*}[htbp]
\centering
\caption{Popularity and difficulty of Logging topics.}
\label{tab:popular_difficulty}
\begin{adjustbox}{width=\textwidth}
\begin{tabular}{lccccc}
\toprule
\multirow{2}{*}{\textbf{Topic}} & \multicolumn{2}{c}{\textbf{Popularity Metrics}} & \multicolumn{3}{c}{\textbf{Difficulty Metrics}} \\
\cmidrule(lr){2-3} \cmidrule(lr){4-6}
 & \textbf{Avg. Views} & \textbf{Avg. Scores} & \textbf{No Answers \%} & \textbf{No Acc. Answers \%} & \textbf{Median Hours to Acc. Answer} \\
\midrule
Custom Logging Frameworks & 3195.12 & 2.90 & 16.42 & 51.65 & 29.51 \\
Data and Tensorboard Logging & 2693.67 & 2.15 & 19.48 & 59.06 & 22.00 \\
Error Handling and Debugging & 3425.10 & 2.05 & 20.43 & 58.67 & 23.83 \\
Event Logging and Monitoring & 2948.57 & 1.85 & 18.15 & 56.33 & 31.16 \\
File Logging and Configuration & 3898.75 & 2.37 & 16.38 & 57.00 & 25.54 \\
General Logging Practices & 3196.82 & 2.79 & 14.50 & 51.56 & 21.36 \\
Logback and Application Configuration & 4249.74 & 2.56 & 19.76 & 59.65 & \textbf{118.14} \\
Logging Levels and Output & \textbf{4825.73} & \textbf{4.13} & 15.49 & 52.34 & 56.83 \\
Logging in Containerized Environments & 3985.13 & 2.68 & \textbf{22.96} & \textbf{64.91} & 47.32 \\
NLog and .NET Logging & 3365.45 & 2.95 & 14.57 & 50.35 & 74.89 \\
Testing and CI Pipelines & 3272.13 & 2.91 & 17.35 & 59.92 & 48.54 \\
\midrule
\textbf{Average for all Logging topics} & 3550.56 & 2.67 & 17.77 & 56.49 & 45.37 \\
\bottomrule
\end{tabular}
\end{adjustbox}
\end{table*}

\subsubsection{Survey Validation of Topic Difficulty and Explanations}
\begin{figure}
    \centering
    \includegraphics[width=\textwidth]{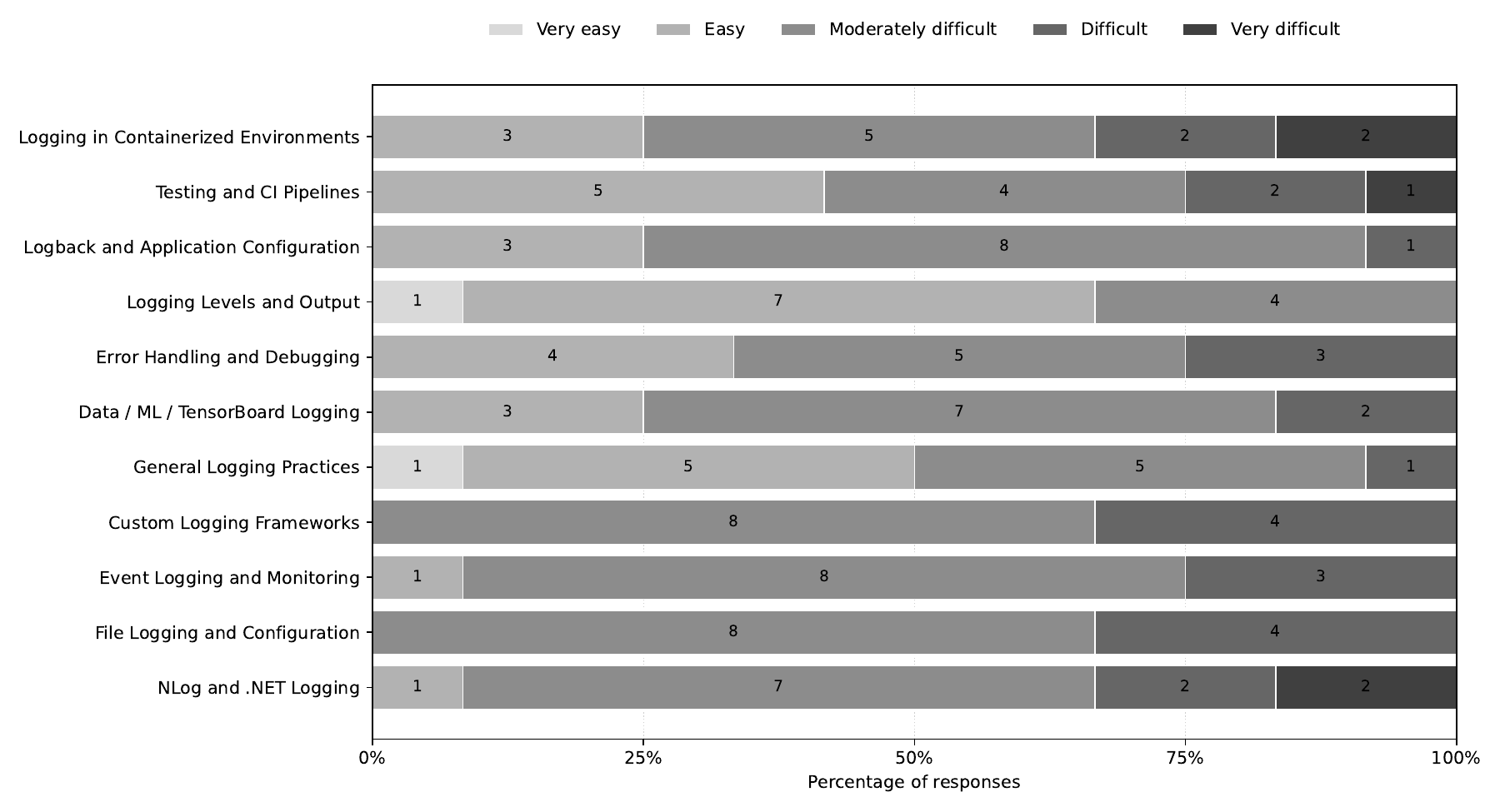}
    \caption{Distribution of survey responses on the perceived difficulty of logging topics.}
    \label{fig:survey_difficulty}
\end{figure}

To complement the quantitative difficulty analysis derived from Stack Overflow (RQ3), we asked practitioners to assess the \emph{perceived difficulty} of each logging topic using a five-point Likert scale ranging from \emph{Very easy} to \emph{Very difficult}. Figure~\ref{fig:survey_difficulty} reports the distribution of responses across topics.

Overall, the survey results are consistent with the Stack Overflow--based difficulty signals. Topics that were empirically hard to resolve in the Q\&A corpus--such as \textit{Logging in Containerized Environments} and \textit{Testing and CI Pipelines}--were also rated as at leat \emph{Moderately difficult} and \emph{Very difficult} by a majority of respondents. For instance, \textit{Logging in Containerized Environments} received 2 \emph{Very difficult} ratings, 2 \emph{Difficult} ratings, and 5 \emph{Moderately difficult} ratings, reflecting the substantial cognitive and operational complexity associated with distributed execution, ephemeral containers, and indirect log routing.
Practitioner justifications help explain these ratings. One respondent emphasized that difficulty sharply increases when logging goes beyond framework defaults and requires custom infrastructure:
\includegraphics[width=2ex]{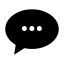} \textit{``Only implementing a custom logging/tracing infrastructure is tricky, as we should consider many aspects of production impacts (e.g., time overhead, storage usage, extensibility, collected info, etc.).''} -- [P1]
This aligns with our RQ3 observation that containerized logging questions often remain unresolved due to environment-specific constraints rather than API misuse.
Similarly, \textit{Event Logging and Monitoring} and \textit{Custom Logging Frameworks} were predominantly rated as \emph{Moderately difficult} to \emph{Difficult}. Respondents noted that, while basic logging is well supported by frameworks, maintaining end-to-end traceability in large-scale or event-driven systems remains challenging:
\includegraphics[width=2ex]{icons.png} \textit{``Most logging frameworks take care of intermediate steps. However, especially in event driven systems, on large scale tracking the log chain is extremely difficult and confusing which requires custom logging.''} -- [P3]
By contrast, foundational topics such as \textit{General Logging Practices} and \textit{Logging Levels and Output} were mostly rated as \emph{Easy} or \emph{Moderately difficult}. This mirrors their lower difficulty in Stack Overflow metrics and reflects the maturity of tooling and shared conventions in these areas. However, even here, respondents highlighted non-technical challenges, such as enforcing consistent practices across teams remain a source of friction:
\includegraphics[width=2ex]{icons.png} \textit{``Ensuring living practice across the team is hard due to lack of rights and wrongs.''} -- [P4]
\includegraphics[width=2ex]{icons.png} \textit{``Unfamiliarity with the process and languages makes it difficult.''} -- [P6]\\
Finally, \textit{Data / ML / TensorBoard Logging} shows an intermediate difficulty profile, with ratings spanning from \emph{Easy} to \emph{Difficult}. Practitioners attributed this variance to the expected level of detail and uneven tooling support in ML pipelines:
\includegraphics[width=2ex]{icons.png} \textit{``The level of details expected to be provided, and also the current tooling support for the types of logging.''} -- [P2]
\includegraphics[width=2ex]{icons.png} \textit{``Logging in  ML system sounds to be different with not standard practice and those ML system require container environment like Docker.''} -- [P8]
\includegraphics[width=2ex]{icons.png} \textit{``logging is sometimes a part of frameworks in machine learning''} -- [P11]
This reinforces our earlier finding that ML logging, while less popular, introduces distinct challenges related to experiment tracking, metric volume, and storage costs.\\
Taken together, the survey confirms that logging difficulty is highly topic-dependent. Difficulty increases when logging intersects with distributed execution, custom infrastructure, or weakly standardized practices--precisely the conditions under which Stack Overflow questions also become harder to resolve.
\subsubsection{Comparative Effort: Logging vs. Other Engineering Tasks}
\begin{figure}
    \centering
    \includegraphics[width=\textwidth]{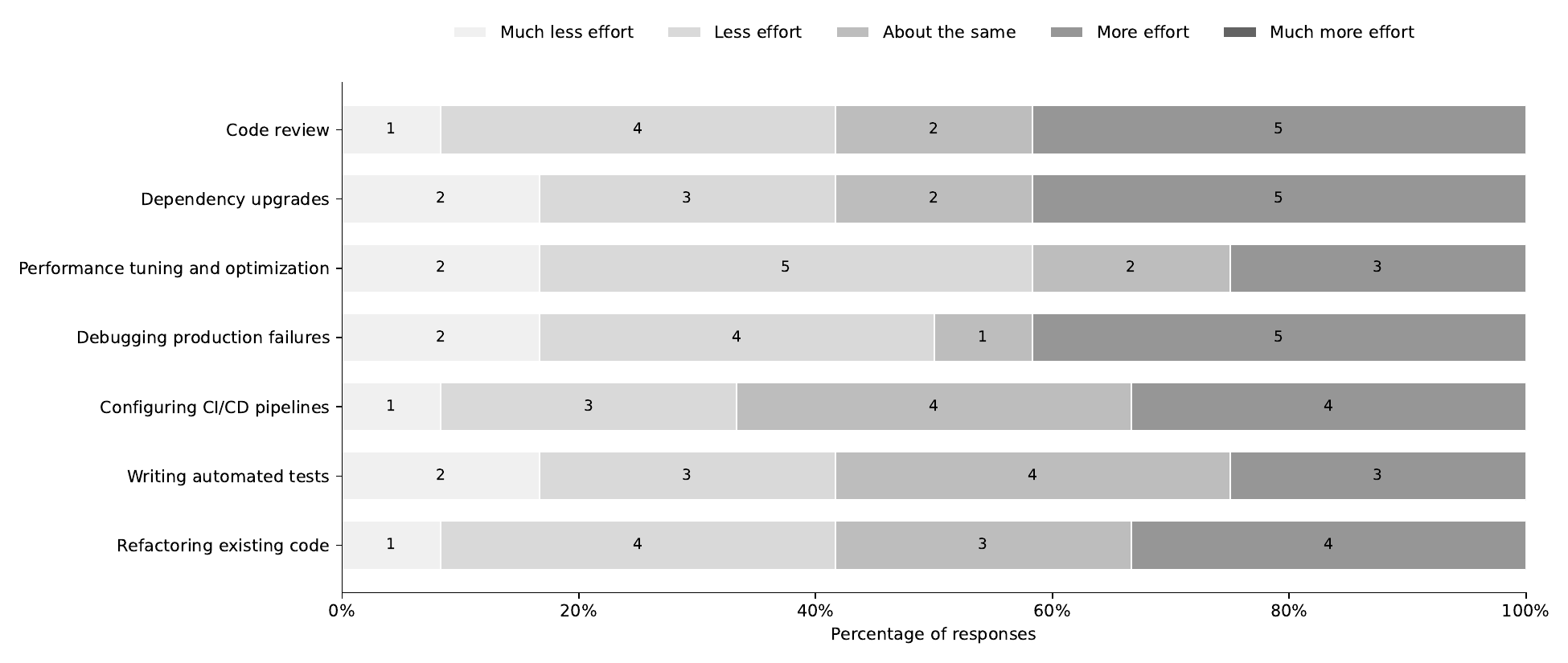}
    \caption{Distribution of survey responses comparing the engineering effort required for logging tasks relative to other common software engineering activities (e.g., code review, refactoring, CI/CD configuration). Higher ratings indicate that logging requires more effort than the corresponding task.}
    \label{fig:survey_effort}
\end{figure}

To contextualize logging effort within broader software engineering activities, we asked participants to compare the effort required for logging-related tasks against other common development tasks, using a five-point scale from \emph{Much less effort} to \emph{Much more effort}. Figure~\ref{fig:survey_effort} summarizes the responses.\\
Overall, respondents generally perceived logging as requiring \emph{about the same} or, in some ontexts, \emph{less effort} than activities such as refactoring, dependency upgrades, or performance tuning. Several participants emphasized that implementing basic logging is often not viewed as the primary engineering challenge within a project:
\includegraphics[width=2ex]{icons.png} \textit{``Basically, implementing logging isn't normally a super sophisticated task. Other options indicated above are much harder and time-taking compared to logging implementations.''} -- [P1]
\includegraphics[width=2ex]{icons.png} \textit{``Most of those task include logging.''} -- [P8]
\includegraphics[width=2ex]{icons.png} \textit{``Logging can be considered a relatively easy task''} -- [P10]\\
However, the comparison reveals important nuances. Tasks such as \textit{Debugging production failures}, \textit{Configuring CI/CD pipelines}, \textit{Code review}, and \textit{Dependency upgrades} were frequently rated as requiring \emph{more effort} than logging, supporting the view that logging acts primarily as an enabling mechanism rather than a primary workload driver. At the same time, respondents noted that logging effort scales with system evolution:
\includegraphics[width=2ex]{icons.png} \textit{``Refactoring requires refactoring the logging system as well.''} -- [P3]
\includegraphics[width=2ex]{icons.png} \textit{``reading codes takes much efforts.''} -- [P11]
Interestingly, logging was perceived as comparable in effort to \textit{writing automated tests}, suggesting that effective logging requires a similar depth of system understanding:
\includegraphics[width=2ex]{icons.png} \textit{``Logging requires as much understanding of the code to write tests, whereas setting up CI/CD pipelines are usually a replicable task using existing information from the internet.''} -- [P4]
Finally, respondents highlighted that effort increases when logging intersects with cross-cutting concerns such as security, cost, and storage management:
\includegraphics[width=2ex]{icons.png} \textit{``Cost of tools and storage, security challenges.''} -- [P11]\\
In summary, while logging is not generally perceived as more effort-intensive than other core engineering activities, its effort profile is highly context-dependent and should be interpreted as indicative practitioners perceptions rather than an objective measurement of effort. The results suggest that logging effort is highly context-dependent and influenced by factors such as system architecture, deployment environment, tooling ecosystem, and operational requirements. Rather than acting as an isolated engineering activity, logging is often intertwined with broader development, debugging, monitoring, and maintenance practices.
\begin{tcolorbox}[colback=black!10!white,colframe=black]
\textbf{Findings:}
\textit{Logging in Containerized Environments} emerges as the most difficult logging topic, exhibiting both the highest proportion of unresolved Stack Overflow questions (64.91\%) and the strongest perceived difficulty in the practitioner survey. In contrast, \textit{Logging Levels and Output} stands out as both the most popular and among the least difficult topics, benefiting from mature tooling and shared conventions. 
Survey results further reveal that logging effort is generally perceived as comparable to or lower than that of other core engineering tasks; however, effort increases substantially when logging intersects with distributed architectures, CI/CD pipelines, and ML workflows. Compared across domains, logging remains more difficult and time-consuming on average based on Stack Overflow difficulty metrics (45.37 hours to resolution) than CI (11 hours), while refactoring appears least difficult overall (36.58\% unresolved; 0.39 hours to resolution).
\end{tcolorbox}

\section{Discussion and implication}
\label{sec: discussion_implication}

We describe the implications of our findings for practitioners, researchers, framework vendors, and educators aiming to improve logging frameworks and practices.

\textbf{Implication for  Researchers.} The results of our study highlight the significant attention that logging practices are receiving across various technology domains, revealing a wide spectrum of challenges that developers encounter, as detailed in our RQ3 findings. These insights present urgent areas for research:
\begin{itemize}
    \item Containerized Environments: Our findings underscore that logging within containerized environments is the most challenging topic in logging discussions on Stack Overflow, as evidenced in Figure \ref{fig:bubble}. We found that this area has received some, but still limited, attention from the research community. Early effort to bridge this gap includes the work of Moilanen et al. \citep{moilanen2020collecting}, who developed a solution for collecting logs from Docker containers, and Chen et al. \citep{chen2020docker}, who explored Docker container log collection using the ELK stack (Elasticsearch, Logstash, and Kibana). Alves et al. \citep{alves2021identifying} also contributed by investigating logging practices in open-source Python containerized applications. However, despite the advancements, several key shortcomings persist. One of the primary challenges remains the ephemeral nature of containers, which often leads to log data being lost when containers are terminated, despite efforts to implement persistent logging solutions \citep{mezmo}. Additionally, while many solutions exist, such as the ELK stack, they often introduce significant overhead, which becomes problematic at scale \citep{chen2020docker}. The integration of these solutions with modern container orchestrators like Kubernetes \citep{eks}, Google Cloud \citep{Googlecloud}, Docker \citep{docker}, and AWS \citep{AWS} remains unexplored. Moreover, while existing research has made important strides, it primarily focuses on specific frameworks or tools (e.g., Docker, Kubernetes, or Python), leaving gaps in understanding for newer or hybrid technologies, such as multi-cloud setups, serverless containers (e.g., AWS Fargate), and microservices architectures transitioning from monolithic systems. Finally, security and isolation of logs, particularly in multi-tenant environments, remain a critical challenge that has not been fully addressed in current solutions \citep{mezmo}. These gaps present opportunities for future research and improvements in containerized logging practices.
    \item Broadening the scope to other fields: Logging is integral to many well-established fields such as machine learning, databases, DevOps, and security, as shown in Figure \ref{fig:tags_concepts}. However, these critical areas receive limited attention from academia, impacting the effectiveness with which practitioners manage logging challenges. Evidence from Table \ref{tab:popular_difficulty} show that  “Testing and CI Pipelines”  exhibits a high unresolved rate (59.92\%) and the second-longest median time to an accepted answer (48.54 h), while typical CI questions resolve far faster (≈ 11 h in \cite{ouni2023empirical}). This disparity indicates that when CI problems involve logging, they are materially harder. We therefore encourage future work on \texttt{logging for Testing \& CI Pipelines} to be situated within the CI topical taxonomy identified by prior CI research--e.g., Build, Version Control, Deployment, Testing, and Culture \cite{ouni2023empirical}--and to examine the role of logging within each topic. Complementarily, researchers should investigate how effective logging mitigates risks in database management and prevents data leaks in ML-intensive systems. These studies could provide new insights that directly impact and improve logging practices across these various domains.
\end{itemize}

\textbf{Implication for Practitioners.} 
Our study clearly indicates which logging topics are most challenging within the development community, guiding team leaders in effectively allocating resources and tasks among their project teams. Based on our findings, areas such as \textit{Logging in Containerized Environments}, \textit{Logback and Application Configuration}, \textit{Testing and CI Pipelines}, \textit{Data and TensorBoard Logging}, and \textit{Error Handling and Debugging} have proven to be particularly complex, as shown in \autoref{fig:bubble}. These topics consistently exhibit greater difficulty, reflected by high proportions of questions without accepted answers and longer median response times.
Because logging is typically embedded within feature development and bug fixing rather than treated as a standalone task, our findings suggest that teams should adopt targeted review practices instead of assigning “logging work” in isolation. Practitioner feedback reinforces this observation, emphasizing that logging decisions often lack systematic oversight:
\includegraphics[width=2ex]{icons.png} \textit{``Logging being overlooked and undervalued makes it difficult to exercise it.''} -- [P4]\\
In particular, senior developers should systematically review logging code across files and modules that involve high-difficulty contexts--containerized systems, CI pipelines, software testing, and AI-based applications--to ensure alignment with system architecture, deployment models, and failure modes. Such cross-cutting reviews help detect subtle issues (e.g., missing context propagation, inconsistent log levels, or logs lost at service boundaries) that are difficult for less experienced developers to anticipate.\\
Our results also highlight the importance of establishing and enforcing consistent logging conventions across teams. Practitioners repeatedly stressed that inconsistency in log structure, verbosity, and semantics increases cognitive load during debugging:
\includegraphics[width=2ex]{icons.png} \textit{``Having consistent log templates, log levels, and amount of information to be logged.''} -- [P1]
Without such conventions, logs become harder to interpret--particularly in distributed systems where developers must reason across multiple services and execution contexts.
In addition, practitioners pointed out that difficulties often stem from poor log readability and lack of structure:
\includegraphics[width=2ex]{icons.png} \textit{``Logs are so complicated and usually unstructured text. It is better to make them easy to read and easy to understand.''} -- [P6]
This suggests that teams should prioritize structured logging and shared formatting guidelines, especially in high-difficulty environments where logs are the primary means of diagnosing failures.
Another important implication concerns onboarding and team knowledge. Several practitioners observed that logging practices are particularly difficult for new team members who lack familiarity with existing system-wide patterns:
\includegraphics[width=2ex]{icons.png} \textit{``Maybe it's most difficult for new employees that might not know the overall patterns of the company or team.''} -- [P10]
This highlights the need for teams to document logging conventions explicitly and integrate them into onboarding processes, rather than relying on tacit knowledge.\\
Finally, our results highlight the need to maintain balanced logging coverage across both difficult and easier areas of the system. In high-difficulty contexts, teams should avoid under-logging that creates blind spots during debugging, while also preventing excessive or redundant logs that obscure relevant signals. Conversely, in empirically less difficult areas--such as \textit{General Logging Practices}, \textit{Logging Levels and Output}, and \textit{NLog and .NET Logging}--developers can rely on standardized templates, consistent log-level conventions, and established configuration patterns to achieve sufficient coverage with minimal overhead.\\
Taken together, these findings suggest that effective logging is not merely a technical concern, but a socio-technical practice that depends on consistent conventions, tooling support, review discipline, and developer education. By embedding logging decisions into everyday development and review workflows, teams can concentrate effort where it yields the greatest diagnostic value, improving system observability and maintainability without treating logging as a separate development activity.

\textbf{Implication for Logging Framework Vendors.} 
Our analysis reveals that certain logging frameworks--notably TensorBoard and Logback--consistently appear in the context of challenging discussions, as evidenced by \autoref{fig:bubble}. These frameworks are associated with topics identified as particularly problematic, reflecting a broader need within the developer community for improved tooling, guidance, and usability support. Practitioner feedback from our survey reinforces this observation, emphasizing that current logging tools often fall short when applied to heterogeneous and evolving system architectures:
\includegraphics[width=2ex]{icons.png} \textit{``First, the tooling should be improved, considering the different types of systems. Although people aim to extend the adoption of tools, in some contexts, such an approach does not properly work.''} -- [P2]
\begin{itemize}
    \item \textbf{Enhanced Documentation and Embedded Guidance.}  
    The prevalence of discussions around \textit{Logging Levels and Output} and \textit{General Logging Practices} (Figure~\ref{fig:horizontal_topic}) underscores the need for logging frameworks to go beyond reference-style documentation. Vendors should provide clearer, opinionated guidance on foundational questions such as what to log, where to log, and how much information to include. This need is echoed by practitioners who stress the importance of consistency and clarity:
    \includegraphics[width=2ex]{icons.png} \textit{``Having consistent log templates, log levels, and amount of information to be logged.''} -- [P1]
    Embedding such guidance directly into frameworks--through defaults, examples, and validation rules--could reduce misconfiguration and improve logging quality, particularly for less experienced developers.
    \item \textbf{Support for Customization Without Excessive Complexity.}  
    With 4.26\% of logging discussions focusing on custom logging frameworks, there is clear demand for greater flexibility. However, practitioner feedback suggests that customization often comes at the cost of increased complexity:
    \includegraphics[width=2ex]{icons.png} \textit{``Tracking and chaining logs.''} -- [P3]
    Framework vendors should therefore aim to offer customization mechanisms that preserve end-to-end traceability while minimizing configuration burden. This includes better support for structured message formatting (e.g., automatic inclusion of correlation IDs), flexible configuration formats (e.g., JSON or code-based configuration instead of XML), and smoother interoperability with external observability pipelines and persistence layers.
    \item \textbf{Improved Support for Log Structure and Readability.}  
    Several practitioners emphasized that logs are often difficult to interpret due to unstructured or overly verbose formats:
    \includegraphics[width=2ex]{icons.png} \textit{``Logs are so complicated and usually unstructured text. It is better to make them easy to read and easy to understand.''} -- [P6]
    This feedback suggests that framework vendors should prioritize structured logging as a first-class feature, offering built-in support for schema definition, field naming conventions, and validation to ensure logs remain both machine-processable and human-readable.
    \item \textbf{Domain-Specific Logging Abstractions.}  
    The intersection of logging with specialized domains--including machine learning, databases, CI/CD, and security--presents an opportunity for domain-aware logging frameworks. Practitioner feedback indicates that generic logging abstractions often fail to capture domain-specific requirements:
    \includegraphics[width=2ex]{icons.png} \textit{``Another challenge is educating developers to adopt such practices.''} -- [P2]
    Domain-specific extensions (e.g., ML experiment-aware logging, CI pipeline–aware log context, or security-sensitive logging defaults) could lower adoption barriers and reduce misuse in these contexts.
    \item \textbf{Reducing the Onboarding Barrier Through Better Defaults.}  
    Finally, practitioners noted that logging frameworks can be particularly difficult for new team members who lack familiarity with system-wide conventions:
    \includegraphics[width=2ex]{icons.png} \textit{``Maybe it's most difficult for new employees that might not know the overall patterns of the company or team.''} -- [P10]
    This highlights the importance of providing sensible defaults, configuration validation, and guided setup workflows that encode best practices and reduce reliance on undocumented team knowledge.
\end{itemize}

\textbf{Implications for Educators:} Our study has identified several logging topics that are prominently discussed on Stack Overflow, which should inform the curricula of training programs and educational materials. Topics such as  \textit{``Logging Levels and Output", ``General Logging Practices", ``Error Handling and Debugging",} and \textit{``File Logging and Configuration,"} represent a significant portion of the logging-related queries. These areas have also been the subject of extensive research, reflecting their importance and complexity in software development. As such, they should be covered extensively in educational programs to equip learners with the necessary skills to address common logging issues effectively. This integration of practical query data with academic insights can help ensure that educational materials are both comprehensive and directly relevant to the needs of the industry.

Moreover, special attention should be given to \textit{``Data and Tensorboard Logging"} and \textit{``Testing and CI Pipelines"} which are connected to the dynamic fields of machine learning and software testing. These topics are becoming increasingly important within the logging sphere, reflecting the evolving nature of technology and the need for specialized knowledge in these areas. Despite their growing importance, there has been comparatively less research tackling these specific challenges directly. This gap highlights an opportunity for academic environments to pioneer research and educational content that addresses these cutting-edge topics. Educators are encouraged to integrate these subjects into their teaching materials to prepare students for the challenges they will face in modern software environments \citep{microsoft1}.

\section{Related works}
\label{sec:related_work}

In this section, we introduce and discuss two areas of related works: (i) research done on logging characteristics, and (ii) research done on developers' challenges. 

\subsection{Studies on characterizing logging practice:}
Logging practices are essential across the software development lifecycle, impacting both traditional software engineering and ML/AI applications. Studies \citep{batoun2024literature, he2022empirical, yang2023interview, gu2022logging} show logging's extensive role in capturing execution details and behavior in various environments. Effective logging has been proven to reduce debugging time and improve diagnostics \citep{li2020qualitative, gu2022logging}. Logging also plays a vital role in software testing, facilitating the identification and analysis of test failures \citep{zhang2022studying}.~\citet{gujral2021exploratory} conducted one of the earliest empirical studies of logging-related discussions on Stack Overflow, applying LDA-based topic modeling to approximately 82K posts to identify high-level themes related to programming languages and object-oriented constructs. Their work demonstrated the feasibility of using Q\&A platforms to study logging concerns and revealed broad topical categories (e.g., logging conversion patterns, Android device logging, and database logging, etc.). 
Our work complements and extends this research in several important ways. Specifically, we analyze a substantially larger and more recent dataset (216,094 logging-related questions) and go beyond topic identification by systematically characterizing logging challenges across topics.

In ML/AI, logging adapts to challenges like complex models and large datasets. \citet{foalem2024studying} highlight the development of ML-specific tools such as MLflow and TensorBoard, which enhance functionality to track model performance, manage hyperparameters, and monitor training metrics, meeting the unique needs of ML applications.

While these studies characterize logging practices in specific domains (general-purpose software or ML/AI), they primarily rely on project-level or tool-level data. None of these existing studies systematically investigates how logging challenges emerge and evolve across a large-scale, community-driven knowledge base such as Stack Overflow. Our study addresses this gap by examining more than 200K logging-related discussions spanning multiple ecosystems and technologies. This enables us to capture not only what logging practices exist but also how developers experience and articulate logging difficulties in practice, providing a broader, cross-domain perspective that previous studies lack.

\subsection{Studies on developers'challenges:}
Understanding developer challenges is crucial for advancing software engineering and enhancing tool support. Stack Overflow, a key resource, provides insights into these challenges through user interactions. Research using Stack Overflow data \citep{openja2020analysis, rosen2016mobile} has revealed prevalent issues in software engineering, detailing the questions developers ask and the solutions provided. As machine learning and AI evolve, new challenges emerge. Studies \citep{islam2019developers, ahmad2020systematic, alshangiti2019developing,zhang2019empirical} have identified specific obstacles in developing and updating ML models. Additionally, \citet{johri2018identifying} used topic analysis to track technology and programming trends on Stack Overflow, demonstrating the impact of community-driven knowledge. A similar study has been conducted by \citet{openja2020analysis} to understand the popularity and challenges of modern release engineering topics.
This research provides invaluable insights into the developer experience, guiding the development of tools and practices to improve productivity and problem-solving in software engineering.
Research has also investigated developer challenges in other critical areas of software engineering. ~\citet{peruma2022refactor} analyzed Stack Overflow posts to characterize challenges related to refactoring, showing that developers frequently struggle with tool support, migration issues, and understanding code dependencies. Similarly, ~\citet{ouni2023empirical} studied questions related to Continuous Integration (CI), identifying pain points in build failures, pipeline configuration, and the integration of new tools into CI environments. These works underscore that Stack Overflow serves as a valuable lens for identifying domain-specific challenges across software engineering.  

Although these studies provide invaluable insights into developer challenges, they are typically domain-specific (e.g., ML/AI, mobile, release engineering) or focus on broad software engineering trends. None of them directly addresses logging as a first-class concern. Our work is the first to systematically investigate logging-related developer challenges on Stack Overflow. By doing so, we uncover trends, pain points, and knowledge gaps unique to logging, complementing prior studies while addressing a neglected but critical aspect of software engineering practice.

\section{Threats to validity}
\label{sec:threats_to_validity}
In this section, we discuss the potential threats to the validity of our research
methodology and findings.

\textbf{Selection of Tags:} Our method of identifying relevant discussions through pre-selected tags and keyword-matching mechanisms might introduce biases. This approach may overlook certain posts that discuss relevant issues but do not use the specific tags or keywords we selected. Although we aim to capture a comprehensive range of challenges by focusing on popular and widely used terms, there is a risk of missing discussions that either use less common terminology or discuss emerging issues not yet well-represented by our chosen keywords.

\textbf{Reliance on a Single Data Source:} By relying solely on Stack Overflow as the data source, we may miss insights available in other forums, blogs, or publications where developers discuss challenges in logging. Although Stack Overflow is a rich and diverse platform used by both novice and expert developers, extending our data sources could provide a more rounded view of the challenges faced in this domain.

\textbf{Impact of External Factors on Logging Practices:} Our study, while extensive, does not account for the specific technological, tool-based, or business system contexts within which logging practices are implemented. The variability in logging statements and practices can be profoundly influenced by the underlying technology stack, the tools employed, and the business requirements driving the system designs. As such, the generalizability of our findings may be limited by our dataset's lack of detailed context about the specific environments in which the logging occurs. Future research could enhance the validity of findings by incorporating studies that link logging practices directly to specific types of technology stacks and business applications, providing a richer, more nuanced understanding of the challenges and practices in diverse development environments.

\textbf{Subjectivity in Data Analysis:} A potential threat arises from the inherent subjectivity involved in defining and categorizing topics from Stack Overflow discussions. To mitigate this, we adopted a structured ground-truth construction process aligning with the methodology of the previous study \citep{abbassi2025unveiling, verdet2023exploring}: two PhD researchers with over six years of experience in software engineering collaboratively applied open coding to a 10\% pilot sample of posts. This process iteratively refined the topics until convergence on 11 stable categories. To further ensure robustness, the two authors independently labeled the final 50\% of the dataset, reaching a substantial inter-rater reliability with a Cohen's Kappa score of 0.781. Disagreements were resolved through consensus discussions. This reliability check reduces the risk of individual bias and is consistent with best practices in empirical software engineering research \citep{foalem2024studying, abbassi2025unveiling, foalem2025logging}. Another threat comes from the use of an LLM (GPT-4o-mini) as a “judge” for topic classification. While LLMs can capture semantic nuances beyond traditional approaches, they may also introduce risks of hallucination, prompt sensitivity, or bias inherited from their training data. We mitigated these risks in several ways: (i) we employed a reference-based judging approach \citep{lin2024engineering}, anchoring the LLM's outputs to the manually validated taxonomy of 11 topics; (ii) we set the temperature to 0 to enforce determinism and reproducibility of outputs, following prior recommendations \citep{verdet2023exploring}; and (iii) we explicitly embedded brief descriptions of each topic in the prompt to reduce the likelihood of hallucinated or irrelevant categories \citep{sollenberger2024llm4vv}.


\textbf{Threats to Measurement Interpretation}: 
In RQ3, we characterize the popularity and difficulty of logging topics primarily using Stack Overflow--based indicators, including view count, post score, the proportion of unanswered questions, the absence of accepted answers, and the time required to receive an accepted answer. Although these metrics are widely adopted in empirical software engineering research, they do not exclusively reflect intrinsic technical difficulty. For example, high view counts may indicate broad interest rather than complexity, while unanswered or unaccepted questions may arise from confounding factors such as incomplete problem descriptions, niche technological contexts, limited community visibility, or rapidly evolving tool ecosystems rather than from inherent difficulty alone.
To mitigate these threats, we adopt several strategies. First, we analyze difficulty at the \emph{topic level} rather than at the level of individual posts, thereby reducing sensitivity to low-quality or idiosyncratic questions. Second, we rely on multiple complementary indicators instead of a single metric, allowing us to triangulate different signals of community struggle. Third, we interpret difficulty comparatively across topics rather than in absolute terms.
In addition, we complement the Stack Overflow--based analysis with an industry-oriented practitioner survey. The survey provides an independent perspective on perceived difficulty and engineering effort, allowing us to assess whether topics that appear difficult at the community level are also experienced as challenging in real-world development settings. However, the survey instrument itself has inherent limitations. The perceived difficulty and effort associated with logging may vary substantially depending on contextual factors such as programming languages, frameworks, deployment environments, observability tooling, organizational practices, system scale, team maturity, and prior experience. Although the survey collected contextual information (e.g., programming languages, architectures, team size, and logging stacks), the current analysis does not explicitly control for all these factors. Furthermore, respondents evaluated topics based on short descriptions and representative examples rather than on complete project-specific contexts, which may lead to variability in interpretation. Consequently, the survey findings should be interpreted as indicative practitioner perceptions rather than objective or universally generalizable measurements of logging complexity or engineering effort. Nevertheless, the survey strengthens measurement interpretation by triangulating community-derived signals with qualitative and quantitative feedback from experienced practitioners.

\section{Conclusion and future work}
\label{sec:conclusion}
This paper introduces a comprehensive empirical study focused on understanding the trends, topics, and challenges in logging practices discussed by developers on Stack Overflow (SO). Our findings show that questions about logging frequently involve programming concepts and framework/library. Leveraging advanced LLM techniques, we have categorized the discussions into 11 distinct topics. Notably, the three predominant topics—\textit{``Logging Levels and Output," ``General Logging Practices,"} and \textit{``Error Handling and Debugging"}—accounting for 70\% of all discussions.

In addition, we observe that \textit{Logging in Containerized Environments} emerged as the most difficult topic, with 64.91\% of its questions lacking accepted answers and a long median resolution time.


From our quantitative and qualitative analyses, combining large-scale Stack Overflow mining with an industry-oriented practitioner survey, we distilled several key insights for different stakeholders. In particular, our findings highlight the substantial engineering effort required to support logging in containerized environments, CI pipelines, and testing workflows. Both community-level difficulty signals and practitioner feedback consistently indicate that logging challenges rarely stem from logging APIs in isolation, but instead emerge from cross-cutting interactions among application code, configuration, execution environments, and operational constraints. The practitioner survey complements the Stack Overflow analysis by validating the industrial relevance of the identified topics and by providing explanatory insights into perceived difficulty and engineering effort. 
As future work, we plan to extend this mixed-methods approach through larger-scale industrial studies, including surveys and in-depth interviews with developers across diverse organizational contexts and team sizes. Such studies would enable finer-grained quantification of effort, time-to-fix, and trade-offs associated with logging decisions, as well as the identification of organizational and tooling constraints that are underrepresented in public Q\&A platforms. Ultimately, this line of work will allow us to further refine and strengthen the practical guidelines proposed in this study.

\section*{Declarations}
\label{sec:declaration}
\textbf{Conflicts of interest:} The authors declare that they have no conflicts of interest relevant to the content of this article.

\noindent\textbf{Funding:} Not applicable.

\noindent\textbf{Ethics approval:} This study was approved by Ethics Committee of Polytechnique Montreal (CER-2324-25-D). All participants provided informed consent prior to their participation.

\noindent\textbf{Clinical Trial Number in the manuscript:} Not applicable.

\noindent\textbf{Informed Consent:} All participants provided informed consent before taking part in the survey.

\noindent\textbf{Data availability statement:} The datasets generated during and/or analysed during the current study are available in the [foalem] repository, [\url{https://doi.org/10.6084/m9.figshare.31062553}].

\noindent\textbf{Author Contributions:} 
\begin{itemize}
  \item \textbf{Patrick Loic Foalem:} Conceptualization, Data Curation, Formal Analysis, Investigation, Methodology, Visualization, Writing -- Original Draft, and Writing -- Review \& Editing.
  \item \textbf{Andre Nguimbous:} Formal analysis.
  \item \textbf{Foutse Khomh:} Project Administration, Resources, Supervision, Validation, and Writing -- Review \& Editing.
  \item \textbf{Heng Li:} Supervision and Writing -- Review \& Editing.
  \item \textbf{Ettore Merlo:} Supervision and Writing -- Review \& Editing.
\end{itemize}











\bibliographystyle{plainnat}
\bibliography{Paper}

\end{document}